\documentclass[pra,aps,a4paper,showpacs,floatfix,twocolumn,superscriptaddress]{revtex4-1}
\usepackage{amsmath}
\usepackage{amsfonts}
\usepackage{amssymb}
\usepackage{graphicx}
\usepackage{dcolumn}
\usepackage{bm}

\DeclareMathOperator{\sgn}{sgn}

\begin{document}

\title{Dynamical Friedel oscillations of a Fermi sea}
\author{J.~M.~Zhang}
\affiliation{Fujian Provincial Key Laboratory of Quantum Manipulation and New Energy Materials,
College of Physics and Energy, Fujian Normal University, Fuzhou 350007, China}
\affiliation{Fujian Provincial Collaborative Innovation Center for Optoelectronic Semiconductors and Efficient Devices, Xiamen, 361005, China}

\author{Y. Liu}
\email{liu\_yu@iapcm.ac.cn}
\affiliation{Institute of Applied Physics and Computational Mathematics, Beijing 100088, China}
\affiliation{Software Center for High Performance Numerical Simulation, China Academy of Engineering Physics, Beijing 100088, China}

\begin{abstract}
We study the scenario of quenching an interaction-free Fermi sea on a one-dimensional lattice ring  by suddenly changing the potential of a site. From the point-of-view of the conventional Friedel oscillation, which is a static or equilibrium problem, it is of interest what  temporal and spatial oscillations the local sudden quench will induce. Numerically, the primary observation is that for a generic site,  the local particle density switches between two plateaus periodically in time. Making use of the proximity of the realistic model to an exactly solvable model and employing the {Abel regularization} to assign a definite value to a divergent series, we obtain an analytical formula for the heights of the plateaus, which turns out to be very accurate for sites not too close to the quench site. The unexpected relevance and the incredible accuracy of the Abel regularization are yet to be understood. Eventually, when the contribution of the defect mode is also taken into account, the plateaus for those sites close to or on the quench site can also be accurately predicted. We have also studied the infinite lattice case. In this case, ensuing the quench, the out-going wave fronts leave behind a stable density oscillation pattern. Because of some interesting single-particle property, this dynamically generated Friedel oscillation differs from its conventional static counterpart only by the defect mode.
\end{abstract}


\maketitle

\section{Introduction}
Quantum nonequilibrium dynamics is now under extensive study  \cite{nonequilibrium,nonequilibrium2,nonequilibrium3}, as theorists are quenching all kinds of models in all kinds of protocols. As a fair observation, in most cases, the model taken up is a many-body system with interaction. Although models of this kind have to be the choice for addressing problems like thermalization, their actual wide employment might to some extent result from the prejudice that only interacting many-body systems could yield novel, interesting dynamics. But actually, when it comes to dynamics, even the simplest model can be nontrivial \cite{akulin,tanor}. The point is that, there is a big gap from the static property of a system to its dynamics, as the extra dimension of time makes the latter much more richer and more complicated than the former. As examples of interesting dynamics exhibited by simple models, we have the Landau-Zener-Stueckelberg tunneling of a two-level system \cite{landau,zener,stueckelberg}, the nonspreading wave packet of a free particle \cite{berry, berry2}, the dynamical localization of a particle on a one-dimensional tight binding chain \cite{dl}, the coherent destruction of tunneling of a particle in a double-well potential \cite{cdt}, etc. Moreover, in the field of atomic physics,  many intriguing  phenomena, such as above-threshold ionization \cite{ati,atiH} and stability under super-intense lasers \cite{eberly}, can be displayed by the simplest atom, namely, the hydrogen atom. Hence, in search of interesting dynamics, one does not need to resort to complicated many-particle systems---single-particle models or many-particle models without interaction could suffice.

\begin{figure}[tb]
\centering
\includegraphics[width= 0.4 \textwidth ]{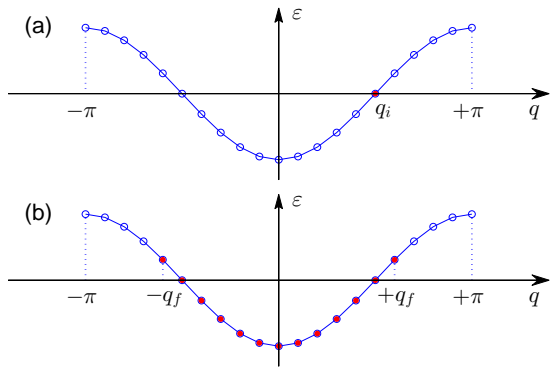}
\caption{(Color online) Two quench scenarios on the one-dimensional tight binding ring. (a) The single-particle case. Initially an arbitrary Bloch state  with wave vector $q_i$ is occupied. (b) The many-particle case. Initially all the Bloch states with wave vector $|q|\leq q_f$ are occupied.
\label{illustration}}
\end{figure}

In this spirit, we had studied the transition dynamics of a Bloch state on a one-dimensional tight binding chain \cite{cusps,plateaus}. The scenario is extremely simple. We just take a one-dimensional tight binding chain with the periodic boundary condition and put a particle in an arbitrary Bloch state [see Fig.~\ref{illustration}(a)]. Then we quench it by suddenly changing the potential of an arbitrary site. Because of the newly introduced local barrier, the particle will be scattered into all other Bloch states, in particular, the one which is the mirror of the initial Bloch state, and its wave function in the real space will be deformed. This is all one can anticipate from general principles. However, down-to-earth numerical simulation reveals that the probability of the particle remaining in its initial state (the survival probability) and the probability of it being reflected into the mirror state (the reflection probability) both show cusps periodically in time [see Fig.~\ref{fig_bloch}(a)]. This kind of nonsmooth behavior is totally unexpected. Yet  more is in store. In the real space, the evolution trajectory of the probability density on a generic site shows plateau structures [see Fig.~\ref{fig_bloch}(b)]. All these unexpected, structured dynamical phenomena can be explained by identifying and studying an ideal model, which we shall sketch below.

It is natural to generalize the single-particle problem to a many-particle one. After all, to single out a particle and prepare it in a certain Bloch state other than the ground one, is no easy task. Experimentally, people more often deal with a collection of  particles simultaneously. Hence, in this paper we consider the same scenario as before but with the initial state being a Fermi sea, in which all the Bloch states below the Fermi energy are filled [see Fig.~\ref{illustration}(b)]. From the point of view of Friedel oscillation, this is a violent way of introducing the defect into the Fermi sea. It is of interest what temporal and spatial oscillation it will induce.  To make things tractable and to adhere to our idea mentioned above, we assume that there is no interaction between the particles. Despite this simplicity, the problem is still very challenging analytically. But at least numerically, it is simple. As there is no interaction, we can just evolve each particle independently, and then for calculating single-particle quantities like the particle number at a site, we just need to sum up the contributions of each particle. It turns out that the particle number at a generic site switches between plateaus of two different heights, with the times of jumping determined by the Fermi wave vector $q_f$.  Determining the heights of the plateaus is then our primary aim.

In the following, we shall first describe in detail the quench scenario and the numerically observed plateaus in Sec.~\ref{sec_scenario}. Then in Sec.~\ref{sec_explanation}, we shall try to develop an analytic formula for the plateaus. This will be done in three steps. First, we shall review the single-particle case in Sec.~\ref{subsec_ideal}, and derive many results in a new way. Then, in Sec.~\ref{subsec_abel}, by approximating the realistic Fermi sea by a fictitious one, we employ the Abel summation trick to get an analytic formula for the heights of the plateaus, which turns out to be very accurate for sites not too close to the quench site. Motivated by this success, we do some formal extension in Sec.~\ref{subsec_formal}, and obtain another analytic formula which reduces to the Abel one for sites far away from the quench site and has the advantage that it is defined also for the quench site. Finally, in Sec.~\ref{subsec_defect}, we take into account the contribution of the defect mode, and get a formula accurate for all sites, including the quench site. This is the ultimate result for the plateau heights. On the basis of this result, in Sec.~\ref{sec_fo}, we consider the infinite lattice case. In this case, there is no repeated plateau switching, and for each site, the particle number will settle down to a constant value. The picture is that the sudden quench generates two out-going wave fronts, which leave in their wakes a stable density pattern. This is Friedel oscillation generated dynamically. We shall discuss its relation with the conventional static Friedel oscillation.

\section{The scenario and the plateaus}\label{sec_scenario}

The setting is simply the one dimensional tight binding chain with the periodic boundary condition, or a lattice ring. The unperturbed single particle Hamiltonian is ($\hbar = 1$ in this paper)
\begin{eqnarray}
  H_0 &=& - \sum_{l=-L}^{L } (|l\rangle \langle l +1 | + |l+1 \rangle \langle l | ) ,
\end{eqnarray}
where $|l\rangle $ denotes the basis function (the Wannier function) at site $l $. Here for notational simplicity, it is assumed without loss of generality that the lattice size $N = 2L +1$ is an odd number. The periodic boundary condition means that $| l \rangle \equiv |l+N \rangle $ for arbitrary $l $. As is well known, the unperturbed eigenstates of $H_0 $ are the so-called Bloch states, which are essentially plane waves on the lattice ring. Explicitly, they have the expression
\begin{eqnarray}
 \langle l | k \rangle  = \frac{1}{\sqrt{N}} \exp(i q  l).
\end{eqnarray}
Here $k$ is an integer indexing the Bloch state and $q = 2\pi k /N $ is the corresponding wave vector. Apparently, $|k \rangle \equiv | k + N \rangle $. It is easily verified that the eigenenergy associated with the Bloch state $|k\rangle $ is $\varepsilon(q) = -2 \cos q $. This dispersion relation is illustrated in Fig.~\ref{illustration}.

In Refs.~\cite{cusps, plateaus}, the following quench scenario [see Fig.~\ref{illustration}(a)] was studied. Initially, a particle is in some Bloch state $|k_i \rangle $ with wave vector $q_i = 2 \pi k_i /N $. This is an eigenstate of $H_0$, and nothing interesting happens. Then at $t=0 $, the system is quenched suddenly by changing the potential of some site, which, because all the sites in a ring are on equal footing, can be assumed to be the $l =0 $ site. The subsequent evolution of the single-particle system is then controlled by the final Hamiltonian $H_f = H_0 + H_1$, with the perturbation
\begin{eqnarray}
  H_1 &=& U |0 \rangle \langle 0 | ,
\end{eqnarray}
where $U $ is the strength of the defect potential, or the quench amplitude.

\begin{figure}[tb]
\centering
\includegraphics[width= 0.45 \textwidth ]{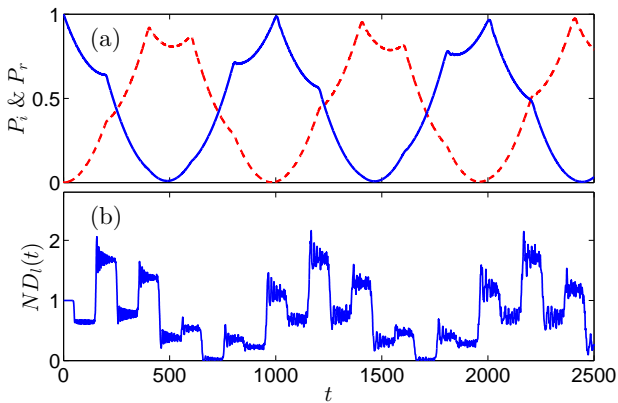}
\caption{(Color online) Quench dynamics of a Bloch state $|k_i \rangle $. (a) Time evolution of the survival probability $P_i$ [solid line, see Eq.~(\ref{pi})] and the reflection probability $P_r$ [dashed line, see Eq.~(\ref{pr})]. (b) Time evolution of the local probability density $D_l $ [see Eq.~(\ref{dl})]. The parameters are $N = 401$, $k_i = 100 $, $l = 100$, and $U = 1.5$. Note the cusps in panel (a) and the plateaus in panel (b). That the survival probability shows cusps has also been observed in some other models \cite{stey,parker,meystre, fermi, scienceopen,ligare,zhou}.
\label{fig_bloch}}
\end{figure}

It turns out that this scenario yields very interesting and unexpected dynamics. Let us denote the wave function of the particle as $|k_i(t)\rangle \equiv \exp(-i H_f t )|k_i \rangle  $, where $k_i$ denotes the initial state and the argument $t$ indicates how long  it has been evolved by $H_f$. This way of notation is used throughout the paper. The state $|k_i (t)\rangle $ can be easily determined once the Hamiltonian $H_f$ is diagonalized numerically. It is then observed that, in the momentum space, both the probability of the particle remaining in its initial state $|k_i\rangle $,
\begin{eqnarray}\label{pi}
  P_i &=& |\langle + k_i | k_i (t)\rangle |^2 ,
\end{eqnarray}
and the probability of it being reflected into the momentum-reversed Bloch state $|-k_i \rangle $,
\begin{eqnarray}\label{pr}
  P_r &=& |\langle -k_i | k_i (t) \rangle |^2 ,
\end{eqnarray}
show cusps periodically; while in the real space,  the local probability density,
\begin{eqnarray}\label{dl}
D_l(t) \equiv |\langle l | k_i (t)\rangle |^2 ,
\end{eqnarray}
at a generic site $l$, jumps constantly from plateau to plateau. These two phenomena are illustrated in Fig.~\ref{fig_bloch}(a) and \ref{fig_bloch}(b), respectively. An important question is at what times the cusps or the sudden jumps occur. The answer is that the cusps are located at $t = r T$, where $r$ is an integer and $T = N/v_i $, with $v_i = 2 \sin q_i $ being the group velocity of a wave packet with wave vector $q_i $, and the sudden jumps are located at $t = r T \pm  s_c $, with $s_c = |l|/v_i $. These results are predicted by the ideal model in Refs.~\cite{cusps, plateaus} and agree with numerics very well.

In the present paper, we study the scenario of locally quenching a Fermi sea, as illustrated in Fig.~\ref{illustration}(b). Now instead of a single Bloch state, all the Bloch states with wave vector $ |q | \leq q_f = 2 \pi k_f /N $ are occupied by spinless, noninteracting fermions. The total number of particles is then $M = 2 k_f + 1$, and the particle number at each site is $\bar{n} = M/N$. The total wave function of the system is always a Slater determinant. Formally, we can denote it as
\begin{eqnarray}\label{totalwavef}
  \Psi(t) &=& Slater \left \{ |-k_f (t)\rangle , \ldots, |k_f(t) \rangle  \right \}.
\end{eqnarray}
Before proceeding, let us introduce another orthonormal basis. Both $H_0 $ and $H_f$ are invariant under the reflection with respect to the defect site $l = 0 $. Hence, it is convenient to have a set of basis states with definite parities. We thus form the even-parity standing waves
\begin{eqnarray}\label{evenstates}
  |k^{+}\rangle  &\equiv & \begin{cases} \frac{1}{\sqrt{2}}(|k\rangle + | -k \rangle ) , & 1 \leq k \leq L , \\ |0 \rangle , & k =0 ,   \end{cases}
\end{eqnarray}
and the odd-parity standing waves
\begin{eqnarray}
  |k^{-}\rangle  &\equiv &  \frac{1}{\sqrt{2}}(|k\rangle - | -k \rangle ) , \quad  1 \leq k \leq  L .
\end{eqnarray}
The odd-parity standing waves vanish at the quench site $l=0$, and thus are not affected by the defect potential. Therefore, they are still eigenstates of $H_f$. But the even-parity standing waves are coupled to each other by the perturbation Hamiltonian $H_1$. In terms of the new basis, the total wave function (\ref{totalwavef}) can be rewritten as
\begin{eqnarray}\label{totalwavef2}
  \Psi(t) &=& Slater \{ |0^{+} (t)\rangle , \ldots, |k_f^{+}(t) \rangle,  |1^{-} (t)\rangle , \ldots, |k_f^{-}(t) \rangle \}. \nonumber
\end{eqnarray}
The quantity of most interest is the local particle number on an arbitrary site $l$. It is the sum of the contributions of all the particles,
\begin{eqnarray}\label{nlt}
  n(l,t) &=& \sum_{k=0}^{k_f} |\langle l | k^+(t)\rangle |^2 + \sum_{k=1}^{k_f} |\langle l | k^-(t)\rangle |^2  \nonumber \\
   &=&  \sum_{k=0}^{k_f} |\langle l | k^+(t)\rangle |^2 + \sum_{k=1}^{k_f} |\langle l | k^- \rangle |^2 .
\end{eqnarray}
Here in the second line we have used that fact that the odd-parity standing waves are eigenstates of $H_f$ and thus they only accumulate a global phase in time. The time variation of the local particle number then comes solely from the even-parity states.

\begin{figure}[tb]
\centering
\includegraphics[width= 0.45 \textwidth ]{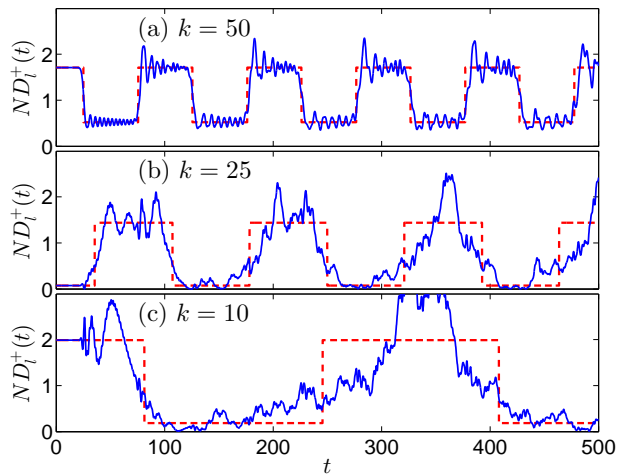}
\caption{(Color online) Time evolution of the local probability density $ D_l^+ (t) \equiv   |\langle l | k^{+} (t)\rangle  |^2 $ with the initial state being an even-parity standing wave $|k^{+}\rangle $. The red dashed lines indicate the analytic prediction (\ref{finalD}) based on the ideal model. In all of the three panels, the common parameters are $N=201$, $l= 50$, and $U = 1.5$.
\label{fig_standing_standing}}
\end{figure}

\begin{figure}[tb]
\centering
\includegraphics[width= 0.45 \textwidth ]{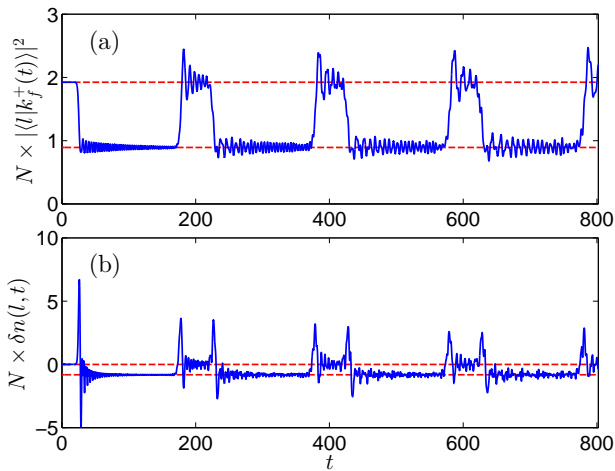}
\caption{(Color online) (a) Time evolution of the local probability density $ |\langle l | k_f^{+} (t)\rangle  |^2 $ with the initial state being the even-parity standing wave $|k_f^{+}\rangle $. (b) Time evolution of variation of the local particle number $\delta n(l,t) = n(l,t) - \bar{n}$, with the initial state being a Fermi sea in which all the Bloch states $|k \rangle $ with $|k|\leq k_f $ are filled. In both panels, the common parameters are $N=401$, $k_f = 100$, $l= 50$, and $U = 1.5$. The dashed lines are guide for the eye, but they are actually the analytic predictions of (\ref{finalD}) or (\ref{abelformula}).
\label{fig_standing_sea}}
\end{figure}

Equation (\ref{nlt}) suggests studying the time evolution of each even-parity standing wave first. Figure~\ref{fig_standing_standing} shows the time evolution of the probability density $D_l^+(t) \equiv |\langle l | k^{+}(t)\rangle |^2$ for three different values of $k$. In Fig.~\ref{fig_standing_standing}(a), we see that $D_l^+ $ alternates between plateaus on two different heights. This should be compared with Fig.~\ref{fig_bloch}(b), where the initial state is not an even-parity state but a Bloch state.  We see that while $D_l^+$ jumps back and forth between just two plateaus, $D_l$ has a sinusoidal profile and shows more plateaus. The difference comes from the fact that $|k (t)\rangle$ contains also an odd-parity part $|k^-(t)\rangle $,  which interferes with the even-parity part $|k^+(t)\rangle $. The remaining two panels of Fig.~\ref{fig_standing_standing} demonstrate a point less emphasized in Refs.~\cite{cusps,plateaus}. There we see that as the wave vector $q = 2 \pi k /N $ recedes from the inflection point $q = \pi/2$, the two-plateau structure deteriorates. This phenomenon has its root in the fact that the dispersion curve $\varepsilon (q)$ is best approximated by a straight line at $q = \pi/2$ (or the middle of the energy band) and least at $q= 0$ or $\pi$ (or the edges of the band). A detailed theory will be presented in Sec.~\ref{subsec_ideal} below.

Now apart from some constant contribution from the odd-parity standing waves, $n(l, t)$ is the superposition of all the curves $|\langle l | k^+(t)\rangle |^2$. In view of the three panels in Fig.~\ref{fig_standing_standing}, it is uncertain what will result from the superposition; in particular, it is uncertain whether $n(l, t)$ will display any plateau structure. Superficially, there are two unfavorable factors. First, for different $k$, the switch times $r T  \pm s_c $ are different, as both $T$ and $s_c$ depend on $k$; second, for those $k\simeq 0 $, the plateau structure is not well developed at all. However, as Fig.~\ref{fig_standing_sea}(b) shows, $n(l, t)$ does show a well-shaped two-plateau structure. Moreover, its switching between the two plateaus is synchronized with that of $|\langle l | k_f^+(t)\rangle |^2$, as is evident by comparing Fig.~\ref{fig_standing_sea}(a) with \ref{fig_standing_sea}(b). Here it is interesting that in $n(l,t)$, the features of the low-lying states are smeared out---only that of the states on the top like $|k_f^+(t) \rangle $ is still visible.
%

\section{Analytic formulas}\label{sec_explanation}

We face the problem to account for the unexpected two-plateau structure displayed by $n(l, t)$. In view of the irregular behavior of the low-lying states, to develop a comprehensive theory seems a task of challenge and is yet to be fulfilled. Here, we confine ourselves to a relatively humble goal. That is, we admit that $n(l, t)$ alternates between two plateaus and that it does so in the same pace as $|\langle l | k_f^+ (t)\rangle |^2$, and focus only on the problem of predicting the heights of the plateaus. Once this is done, the skeleton of the curve of $n(l,t)$ is determined. As the first plateau corresponds to the pre-quench value $\bar{n}$ of $n(l)$,  the real problem is to predict the height of the second plateau.

This is what we try to do in the following subsections. As we shall see, this humble target is still too high, and in many cases we have to resort to bold approximations or just formal procedures to do what we can only do. However, interestingly, the formulas so obtained, although not really rigorously justified, turn out to be very accurate predictions of the height of the second plateau. In particular, in the end, we will get an analytic formula accurate for an arbitrary site $l$.

\subsection{An exactly solvable ideal model}\label{subsec_ideal}

As Eq.~(\ref{nlt}) shows, to calculate $n(l,t)$, we can calculate $D_l^+ (t) = |\langle l | k^+(t)\rangle|^2$ first. In this subsection, we try to find an approximate value of the latter by using the proximity of the realistic model to an ideal model, which is exactly solvable. This has essentially been done in Refs.~\cite{cusps,plateaus} already, so below we just sketch the construction of the ideal model and present the most relevant result, with which the phenomena in Figs.~\ref{fig_bloch}, \ref{fig_standing_standing}, and \ref{fig_standing_sea}(a) can be understood. For the sake of completeness, the detailed derivation is retained but deferred to the Appendix. The derivation there is more straightforward than that in Refs.~\cite{cusps,plateaus}. We shall also discuss more about the limitation or validity of the approximation than in Refs.~\cite{cusps,plateaus}.

The state $|k_i^+(t)\rangle $ evolves by the equation
\begin{eqnarray}
  i \frac{\partial }{\partial t} |k_i^+(t)\rangle  &=& ( H_0 + H_1 ) |k_i^+(t)\rangle ,
\end{eqnarray}
with the initial value $|k_i^+(t=0)\rangle = |k_i^+ \rangle$. As both $H_0$ and $H_1$ are diagonal with respect to the even-odd partition of the basis functions, $|k_i^+(t)\rangle $ always belongs to the even-parity subspace, and is of the form
\begin{eqnarray}
  |k_i^+(t)\rangle  &=& \sum_{k=0}^{L} a_k (t) |k^+\rangle ,
\end{eqnarray}
where $a_k(t)$ are time-dependent coefficients. With respect to the even-parity basis functions $\{ |k^+ \rangle \}$, $H_0$ is diagonal while $H_1$ is close to a matrix whose entries are all equal. Specifically, we have
\begin{eqnarray}
  \langle k_1^+ | H_0 | k_2^+ \rangle  = -2 \cos (2\pi k_1 /N) \delta_{k_1, k_2} ,
\end{eqnarray}
and ($g \equiv  U/N $)
\begin{eqnarray}\label{coupling1}
  \langle k_1^+ | H_1 | k_2^+ \rangle = \begin{cases} 2g , & \text{if both $k_{1,2}$ are  nonzero}  ,\\ \sqrt{2}g , & \text{if only one of $k_{1,2}$ is zero},\;\;\;\; \\ g, & \text{if both $k_{1,2}$ are zero} .     \end{cases}
\end{eqnarray}
Numerically, it is observed that for given $U$, if the lattice size $N$ is large enough, and if the wave vector $q_i $ is not so close to $0$ or $\pi$, in the time evolution, only those few states $|k^+\rangle $ with $k\simeq k_i $ are significantly populated. Other states are far off resonance and hence are suppressed. This fact means that in the full spectrum of $H_0 $, only a small segment is effective in the dynamics in question. Within this segment, the spectrum of $H_0 $ is nearly equally spaced, with the gap between two adjacent levels being approximately
\begin{eqnarray}
  \Delta = \varepsilon(q_i + \eta ) - \varepsilon(q_i )   &\simeq & \eta  \varepsilon'(q_i ) =  \frac{4\pi}{N} \sin q_i  ,
\end{eqnarray}
where $\eta = 2 \pi /N $ is the difference between two adjacent wave vectors in the Brillouin zone. Also, by (\ref{coupling1}),  the coupling between two arbitrary levels in the segment is of the constant value $2g $.

\subsubsection{Construction and solution of the ideal model}\label{idealmodel}

We now extend this segment, together with the two features,  to the whole axis $(-\infty, + \infty )$. We consider an ideal model consisting of infinitely many levels $\{ | \tilde{j} \rangle | - \infty \leq j \leq + \infty  \}$, whose energies are equally spaced, i.e.,
\begin{eqnarray}\label{tildeH0}
  \tilde{H}_0 | \tilde{j } \rangle  &=&  j \Delta  |\tilde{j}\rangle ,
\end{eqnarray}
and between any pair of which, the coupling is constant, i.e.,
 \begin{eqnarray}\label{tildeH1}
   \langle \tilde{j}_1| \tilde{H}_1 | \tilde{j}_2 \rangle  &=& 2 g ,
 \end{eqnarray}
for $ \forall j_1, j_2 \in \mathbb{Z} $. Here $\tilde{H}_{0,1}$ are counter-parts of $H_{0,1}$. By construction, $|\tilde{j}\rangle $ is meant to be the counter-part of $|(k_i + j )^+ \rangle $, and its wave function in the real space is postulated to be
\begin{eqnarray}
  \langle l | \tilde{j}\rangle  = \langle l | (k_i + j )^+\rangle =  \sqrt{\frac{2}{N}} \cos [(q_i + j \eta )l] .
\end{eqnarray}
The correspondence between $|\tilde{j}\rangle $ and $|(k_i + j )^+\rangle $ of course fails for $j $ large enough, either because $|(k_i + j )^+ \rangle$ no longer belongs to the linear segment, or simply because there is no one-to-one correspondence between a finite set and an infinite one. However, as long as those fictitious levels $|\tilde{j}\rangle $ are not significantly populated, the error introduced could be negligible. Also note that in (\ref{tildeH0}), the energy of $|\tilde{j}\rangle $ is defined as $j \Delta$ but not $\varepsilon(q_i ) + j \Delta $. This is a mere shift of the origin of energy, and does not lead to any observable effect.

Corresponding to the original dynamical problem of the realistic model, in the ideal model, initially the system is in the state $|\tilde{\psi} (t=0)\rangle  = | \tilde{0 }\rangle $. It then evolves as
\begin{eqnarray}\label{seq2}
  i \frac{\partial}{\partial t} | \tilde{\psi} \rangle = (\tilde{H}_0 + \tilde{H}_1)| \tilde{\psi}\rangle   .
\end{eqnarray}
At any time $t$, it is of the form $| \tilde{\psi} (t)\rangle  = \sum_{j\in \mathbb{Z} } \tilde{a}_j (t) | \tilde{j} \rangle $.
As $|\tilde{j} \rangle $ is the counterpart of $|(k_i + j )^+ \rangle $, so is $\tilde{a}_j (t)$ that of $a_{k_i + j} (t) e^{i \varepsilon(q_i) t }$. Once $\tilde{a}_j(t)$ is calculated, we get an approximation of $\langle l | k_i^+(t) \rangle $ as
\begin{eqnarray}\label{approx1}
  \langle l | k_i^+ (t) \rangle &=  & \sum_{k=0}^L a_k(t) \langle l | k^+ \rangle \nonumber \\
  & \simeq  & e^{-i \varepsilon(q_i) t } \sqrt{\frac{2}{N}} \sum_{j\in \mathbb{Z}  }  \tilde{a}_j (t)  \cos [(q_i + j \eta )l] .\;\; \;
\end{eqnarray}

\begin{figure}[tb]
\centering
\includegraphics[width= 0.45\textwidth ]{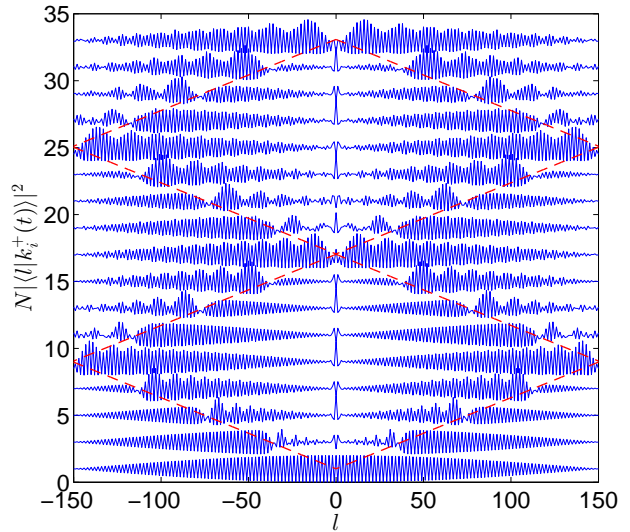}
\caption{(Color online) Equal-distant snapshots of the probability distribution of the wave function $|\langle l | k_i^+(t)\rangle |^2$ in the time interval $[0, 2T]$. For clarity, two adjacent curves are displaced by an amount of 2 in the vertical direction. The parameters are $(N, k_i, U) = (301,75, 2) $. The dashed lines indicate the motion of the wave fronts with a constant velocity of $v_i = 2 \sin q_i \simeq  2 $.
\label{fig_snapshots_single}}
\end{figure}

Now we have two steps to go. First, we have to solve $\tilde{a}_j (t)$ and then we can substitute it into (\ref{approx1}) to calculate $ \langle l | k_i^+ (t) \rangle$ and in turn $D_l^+(t)$  approximately. The detailed calculation is in the Appendix. Here we just describe the ultimate result. For the current problem, a characteristic time scale is the so-called Heisenberg time $T = 2 \pi /\Delta $. For a given site $l$ satisfying $-L \leq l \leq L $, there are two critical times, $s_c^+ = T |l|/N$ and $s_c^- = T (N-|l|)/N$. The physical meaning of these times will be clear in the next paragraph. Now for an arbitrary time $t = r T + s$, where $0<s < T$ and $r$ is a nonnegative integer, we have
\begin{eqnarray}\label{finalD}
  D_l^+(t) &\simeq  & \begin{cases} \frac{2}{N} \cos^2 q_i l , & \text{if } s \in (0 ,s_c^+ )  , \\
  \frac{2}{N} \cos^2 (q_i |l | - \theta_i /2 ) , & \text{if } s \in (s_c^+ ,s_c^- )  , \\
  \frac{2}{N} \cos^2 q_i l , & \text{if } s\in ( s_c^- ,  T ) ,   \end{cases}
\end{eqnarray}
where the angle variable $\theta_i = \theta(q_i ) $ is defined as
\begin{eqnarray}\label{deftheta}
  \theta (q)  = 2 \arctan \frac{U}{2 \sin q }.
\end{eqnarray}

Equation (\ref{finalD}) explains the two-plateau structure in Figs.~\ref{fig_standing_standing} and \ref{fig_standing_sea}(a). Its content can be best illustrated by presenting the snapshots of the probability distribution $|\langle l | k_i^+(t) \rangle |^2$ at consecutive times, as is done in Fig.~\ref{fig_snapshots_single}. There, 17 snapshots of $|\langle l | k_i^+(t) \rangle |^2$, which are sampled equal-distantly in the time interval $[0, 2T]$, are stacked in a pile. The red dashed lines indicate the fronts of the waves generated by the sudden quench. They move at the constant velocity $v_i = 2 \sin q_i $. The time for them to finish a loop on the lattice ring is then $N/v_i$, which is exactly $T $. Similarly, $ r T + s_c^\pm$ are just the times when the wave fronts (backwards or forwards) pass the site $l$.  One can recognize that the curves  show a pattern inside the two rhombuses and another pattern outside the two rhombuses. On the interfaces between the two regions, the dislocation is quite clear. This observation embodies the meaning of  (\ref{finalD}) that $|\langle l | k_i^+(t) \rangle |^2$ alternates between two values in time.

Another thing noticeable in Fig.~\ref{fig_snapshots_single} is that at $t = r T$, the probability distribution $|\langle l | k_i^+(t) \rangle |^2$ reconstructs itself to its initial pattern approximately. It would be an exact reconstruction if it were the ideal model---the ideal model has an equal-distant spectrum of (\ref{spectrum}), hence the revival would be periodic and complete.

Some new light will be shed on (\ref{finalD}) when we consider the eigenstates of $H_0+H_1 $ in Sec.~\ref{sec_fo}.

\subsubsection{Validity of the approximation}\label{limitation}
Now we discuss the validity of the approximation. The ideal model agrees with the realistic model only locally. Hence, for the dynamics occurring in the ideal model to be a faithful reflection of that occurring in the realistic model, the levels actively participating (significantly populated) in the ideal model should be as few as possible, and the linearization of the spectrum $\varepsilon(q)$ at $q = q_i $ should be as good as possible. By (\ref{ajt2}), we see that the population on $|\tilde{j}\rangle $ ($j\neq 0 $) is up bounded by the value
\begin{eqnarray}
  \max_{t} |\tilde{a}_j(t)|^2 &=& F(g/\Delta )/j^2 ,
\end{eqnarray}
where $F(x) \equiv  16 x^2/ (1 + 4 \pi^2 x^2 ) $. As the series $\sum_{j\geq 1 } 1/j^2$ converges absolutely, for any $\epsilon > 0$, there exists a $n_{max}$, such that $\sum_{|j|> n_{max}} |\tilde{a}_j(t)|^2 < \epsilon $. Moreover, as $F(x)$ is a monotonically increasing function of $|x|$, the smaller the ratio $|g/\Delta |$ is, the smaller $n_{max}$ can be chosen. As $g/\Delta = U/(4 \pi \sin q_i )$, for fixed $U$ and $N$, the minimum of $|g/\Delta | $ is achieved at $q_i = \pi/ 2$.

On the other hand, the energy of the state $|(k_i + j )^+ \rangle $ has the Taylor expansion
\begin{eqnarray}\label{taylor}
  \varepsilon(q_i + \eta j ) &=& \varepsilon(q_i ) + \varepsilon'(q_i ) \eta j + \frac{1}{2} \varepsilon''(q_i) \eta^2 j^2 + \ldots .\;\;
\end{eqnarray}
At $q_i = \pi/2$, the quadratic term vanishes identically, and the linearization is the best. Hence, the ideal model as an approximation works the best at $q_i = \pi/ 2$. As $q_i $ recedes from this optimal point (either towards 0 or $\pi$), the ratio $|g/\Delta |$ increases and at the same time the quadratic term in (\ref{taylor}) increases too. Both facts are detrimental to the approximation, and the analytic predictions above become less accurate. This is demonstrated in Fig.~\ref{fig_standing_standing}.

Although $\pi/2$ is the most favorable value of $q_i $, at other levels of $q_i$, the approximation can work well too as long as the lattice size $N$ is large enough. Actually, as $g/\Delta = U/(4\pi \sin q_i )$ is independent of $N$, $n_{max}$ is independent of $N$. However, for $j $ in the range $- n_{max}\leq j \leq n_{max}$, the quadratic term will become negligible in comparison with the linear term as $N\rightarrow \infty $. Therefore, for given $U$, the range of $q_i $ where the ideal model is a good approximation will become larger and larger as $N$ increases, and will cover $(0,\pi)$ eventually in the limit of $N = \infty $.

\subsection{Magic of the Abel summation method}\label{subsec_abel}

\begin{figure}[tb]
\centering
\includegraphics[width= 0.43 \textwidth ]{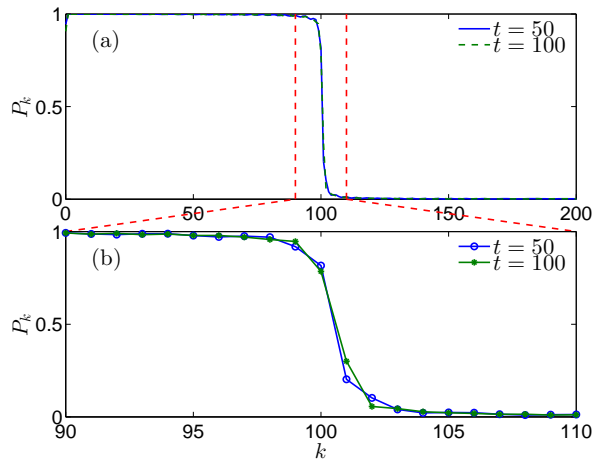}
\caption{(Color online) (a) Population on each even-parity standing wave $|k^+\rangle $ at two arbitrarily chosen times [see Eq.~(\ref{pk}) for the definition]. The parameters are $N = 401 $, $k_f = 100$, and $U =2 $. (b) Highlight of the transition region in (a).
\label{fig_pop}}
\end{figure}

Having treated the single-particle problem, we now proceed to the many-particle problem. By (\ref{nlt}), we face the problem of summing over the $|k^+(t)\rangle $ states. But  the problem is that, as pointed out in the proceeding subsection, for a finite lattice and for states $|k^+\rangle$ lying close to the bottom of the energy band, we do not have an accurate estimation of $|\langle l | k^+(t)\rangle |^2 $.

This seems an insurmountable difficulty. But there is an important fact worth noting. In Fig.~\ref{fig_pop}, we study the population on each even-parity state $|k^+\rangle $ ($0\leq k \leq L $)
\begin{eqnarray}\label{pk}
  P_k(t) &=& \sum_{k_i =0}^{k_f} |\langle k^+ | k_i^+(t)\rangle |^2
\end{eqnarray}
at some arbitrary times. We see that the population distribution is close to its initial value
\begin{eqnarray}
  P_k(t=0) &=& \begin{cases}  1, & \text{if } k \leq k_f , \\ 0, & \text{if } k > k_f .   \end{cases}
\end{eqnarray}
Hence, although on the single-particle level, each $|k^+ \rangle $ state mixes with its neighbors, on the many-particle level, population transfer occurs only between the states in the vicinity of the Fermi energy. Those states at the bottom of the Fermi sea, which pose a difficulty for the summation, are dormant.

Actually, that the states deep beneath the surface of the Fermi sea are  in a sense dormant is also reflected in Fig.~\ref{fig_standing_sea}. Originally, $n(l, t)$ is the superposition of all $|\langle l | k^+(t)\rangle |^2 $ with $0\leq k \leq k_f $. By (\ref{finalD}) and as demonstrated in Fig.~\ref{fig_standing_standing}, these quantities switch between plateaus at different times. However, the net result of the superposition is that $n(l, t)$ switches between the two plateaus with the same rhythm as $|\langle l | k_f^+(t)\rangle |^2 $.

The observations above motivate us to approximate the realistic Fermi sea by a fictitious one. Suppose that in the fictitious model in the proceeding subsection, all the levels $|\tilde{j} \rangle $ with $j \leq 0 $ are filled initially. After the quench, the levels mix with each other, but overall, population transfer occurs only at the surface layer of the Fermi sea. For this fictitious model, the analytic formula (\ref{finalD}) works exactly for all the single-particle levels, and the plateau switchings are all synchronized. {The fictitious Fermi sea is infinitely deep, and hence the total number of particles on a local site $l$ is ill-defined, but its variation could be well defined}. By (\ref{finalD}), the height of the second plateau relative to the first plateau is
\begin{eqnarray}\label{abelseries}
  \delta n(l) &=& \frac{1}{N} \sum_{m\geq 0 }^\infty \bigg\{     \cos \left[ 2 (q_f - \eta m )|l| - \theta_f \right ] \nonumber \\
   & & \quad \quad \quad \quad \quad  -  \cos\left [2 (q_f - \eta m ) |l|  \right ] \bigg\} .
\end{eqnarray}
Here $\theta_f \equiv \theta (q_f )$. Apparently, this series is not convergent but oscillatory. But one can use the trick of the Abel summation method to extract a finite and definite value from it \cite{hardy}. The idea is that while the series
\begin{eqnarray}\label{oriseries}
   \sum_{m=0}^\infty a_m
\end{eqnarray}
diverges, the series
\begin{eqnarray}
 I(x) \equiv   \sum_{m=0}^\infty a_m x^m
\end{eqnarray}
might converge for $0\leq x <1 $, and the limit
\begin{eqnarray}
  \lim_{x\rightarrow 1^-}  I(x) = S
\end{eqnarray}
might exist. If so, then the original summation (\ref{oriseries}) is assigned the value $S$. Carrying out these procedures for the series (\ref{abelseries}), which involve just some geometric series, we get readily
\begin{eqnarray}\label{abelformula}
  \delta n_A(l) &=& \frac{\sin(2 q_f |l| - \theta_f + \eta |l| ) - \sin(2 q_f |l| + \eta |l| )}{2 N \sin (\eta |l|) } \nonumber \\
  &=& - \frac{\sin (\theta_f/2) \cos (2 q_f |l| + \eta |l| - \theta_f /2)}{N \sin (\eta |l|) } .
\end{eqnarray}
Here the subscript indicates that the formula is from the Abel summation method. It turns out that this formula, which is obtained by a purely formal procedure, is a very accurate prediction of the second plateau in the time evolution of $n(l,t)$.

\begin{figure}[tb]
\centering
\includegraphics[width= 0.45 \textwidth ]{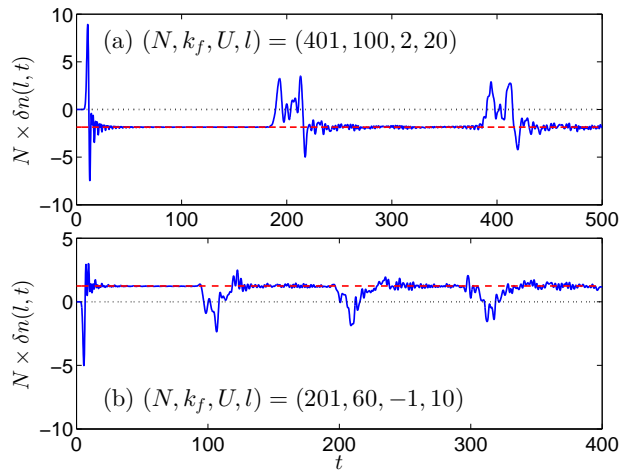}
\caption{(Color online) Time evolution of the variation of the local particle number $\delta n (l,t) \equiv  n(l,t) - \bar{n} $. In each panel, the blue solid line is the numerically exact result, while the horizontal red dashed line is the analytic prediction of Eq.~(\ref{abelformula}). The parameters are displayed in each panel.
\label{fig_time_abel}}
\end{figure}

\begin{figure}[tb]
\centering
\includegraphics[width= 0.45 \textwidth ]{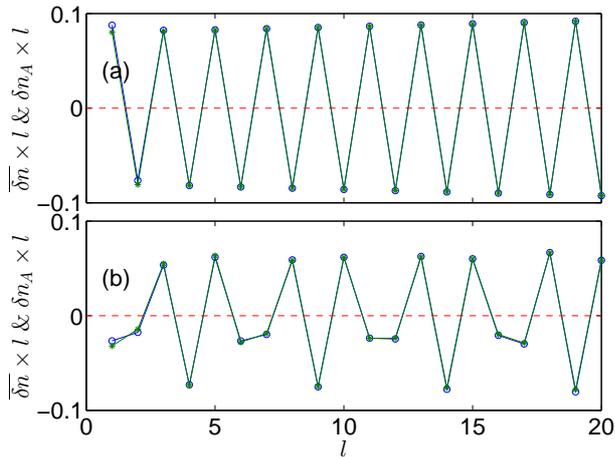}
\caption{(Color online) More systematic check of the accuracy of the analytic formula (\ref{abelformula}) by comparing $\overline{\delta n }$ [see Eq.~(\ref{defp}), the $\circ$ markers] and $\delta n_A $ [see Eq.~(\ref{abelformula}), the $\ast$ markers] for a wide range of $l$. In (a) [(b)], the parameters $(N, k_f, U)$ are of the same values as in Fig.~\ref{fig_time_abel}(a) [Fig.~\ref{fig_time_abel}(b)]. As $\delta n_A(l)$ decays like $1/l$, for clarity, we present $\overline{\delta n} \times l $ and $\delta n_A \times l $.
\label{fig_exact_abel}}
\end{figure}

In Fig.~\ref{fig_time_abel}, we show the evolution trajectories (solid lines) of the variation of local particle number $\delta n (l,t) = n(l,t) - \bar{n} $ for two different sets of parameters. The analytic prediction of (\ref{abelformula}) is indicated by the horizontal dashed lines. In each panel, we see that in the time intervals where the second plateau is expected, the solid line either almost coincides with the horizontal dashed line, or oscillates around it.

In Fig.~\ref{fig_exact_abel}, we check the accuracy of the formula (\ref{abelformula}) in a more systematic way. For the two sets of parameters $(N, k_f, U)$ in Fig.~\ref{fig_time_abel}, we compare the plateaus exhibited by $\delta n (l,t)$ and the analytic prediction $\delta n_A(l)$ in (\ref{abelformula}) for $1 \leq l \leq 20 $. The height of the plateau is defined as
\begin{eqnarray}\label{defp}
  \overline{\delta n} (l ) &=& \frac{1}{t_2 - t_1 } \int_{t_1}^{t_2}  \delta n (l, \tau ) d \tau ,
\end{eqnarray}
where $t_{1,2}$ are chosen in the region where $\delta n $ has settled down. In our calculation, we take $t_1 = 0.5 s_c^+ + 0.5 s_c^- $ and $t_2 = 0.1 s_c^+ +0.9 s_c^-  $. This choice of $t_{1,2}$ is of course arbitrary, but other choices yield almost the same result.
As $\delta n_A (l)$ decays like $1/|l| $, for clarity, instead of $\overline{\delta n }$ or $\delta n_A $, we present $\overline{\delta n } \times l $ and $\delta n_A  \times l $ in Fig.~\ref{fig_exact_abel}. We see that the two agree with each other very well in the whole range of $l$. The difference is only discernible at the lower limit $l=1$. But even there, the relative error is smaller than $20\%$.

\begin{figure}[tb]
\centering
\includegraphics[width= 0.45 \textwidth ]{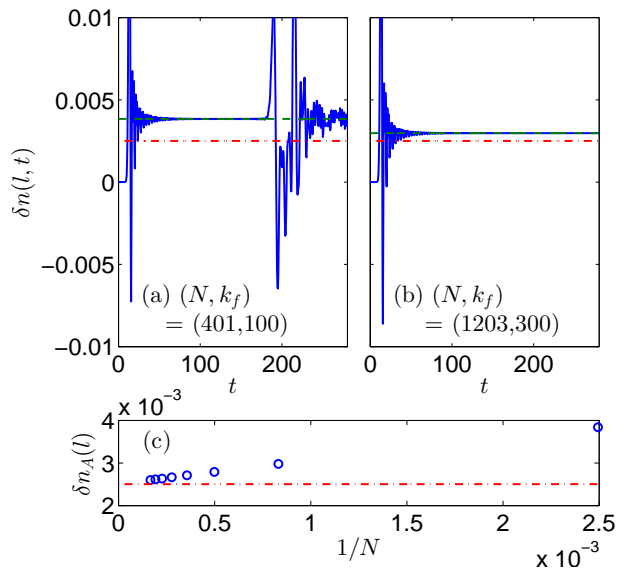}
\caption{(Color online) In the thermodynamic limit, $\delta n_A (l)$ in (\ref{abelformula}) converges to $\delta n_A^\infty (l)$ in (\ref{abellimit}). This is illustrated in (a) and (b) by fixing $(U,l)=(2,25)$, and enlarging $(N,k_f)$ from $(401,100)$ to $(1203,300)$. In each panel, the solid line denotes the numerically exact time-evolution of $\delta n (l,t)$, while the dashed line the Abel formula (\ref{abelformula}), and the dash-dotted line its thermodynamic limit (\ref{abellimit}). A more systematic study is presented in (c). Fixing $(U,l,N/k_f)$ and starting with $(N,k_f) =( 401, 100)$, $(N,k_f)$ are enlarged consecutively by 1, 3, 5, \ldots, 15 times. The circles denote the predictions of $\delta n_A (l)$. Apparently, they fall on a straight line with respect to $1/N$ and, as $1/N \rightarrow 0 $, converge towards $\delta n_A^\infty (l )$, which is indicated by the horizontal line.
\label{fig_converge}}
\end{figure}

The quantity $\delta n_A $ depends on the lattice size $N $. But in the thermodynamic limit of $( N ,k_f ) \rightarrow \infty $ with $q_f =2 \pi  k_f/ N$ fixed, it converges to
\begin{eqnarray}\label{abellimit}
  \delta n_A^\infty (l) &=& \lim_{N\rightarrow \infty } \delta n_A (l ) \nonumber \\
  &=& -  \frac{ \sin (\theta_f/2)  \cos (2 q_f |l| - \theta_f/2 ) }{2\pi |l| }  ,
\end{eqnarray}
which is a cosine function modulated by the $1/|l| $ function. This limiting behavior is illustrated in Fig.~\ref{fig_converge}. In Figs.~\ref{fig_converge}(a) and \ref{fig_converge}(b), we study two cases with the Fermi wave vectors $q_f$ being the same while the lattice sizes $N$ differing by a factor of three. In each panel, the red dash dotted line indicates the value $\delta n_A^\infty $, which is independent of $N$. From Fig.~\ref{fig_converge}(a) to \ref{fig_converge}(b), we see clearly that as the lattice size increases, the plateau, which is accurately predicted by $\delta n_A $, tends towards the dash-dotted line predicted by $\delta n_A^\infty $. This means that the system shows some finite-size effect and that is accurately captured by $\delta n_A $. The scaling behavior of the finite-size effect is investigated more systematically in Fig.~\ref{fig_converge}(c), where the lattice size $N$ in Fig.~\ref{fig_converge}(a) is further increased by 5, 7, \ldots, 15 times, and $\delta n_A $ is plotted against $1/N$.
It is apparent that $\delta n_A $ (the circles) approaches $\delta n_A^\infty $ (the horizontal line) by the $1/N $ power law.

We have thus seen that the formal result (\ref{abelformula}) is a very accurate prediction of the height of the second plateau. Its success is the most dramatic thing in this paper. To make sense of the magic Abel summation method, we note that it is essentially a way of regularization. The exponentially decaying factor $x^m $ (when $0< x < 1 $) helps to suppress the contribution of the low-lying states (with $m \gg 1  $), which are fictitious in the fictitious Fermi sea and dormant in the realistic Fermi sea.

Of course, formula (\ref{abelformula}) also has its drawbacks. In the first place, it is not defined at the quench site $l = 0$. This corresponds to the fact that the series $\sum_{m \geq 0 } 1 $ is not Abel-summable. This shortcoming will be overcome below by other formulas.

\subsection{A formal summation}\label{subsec_formal}
\begin{figure}[tb]
\centering
\includegraphics[width= 0.45 \textwidth ]{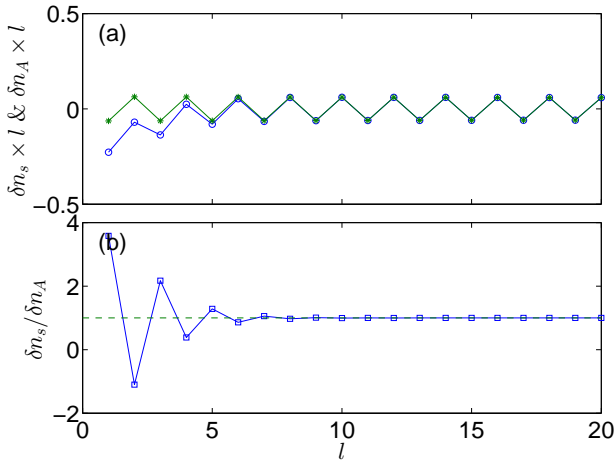}
\caption{(Color online) To verify that in the large $l$ limit, the formal summation (\ref{ns}) reduces to the Abel formula (\ref{abelformula}). The parameters are $(N, k_f, U) = (401, 100, -1)$. In (a), for clarity, $\delta n_s \times l $ ($\circ$ markers) and $\delta n_A \times l $ ($\ast $ markers) are presented. In (b), the horizontal dashed line indicates the value of unity.
\label{fig_abel_sum}}
\end{figure}

The success of the formal procedure in the proceeding subsection is very encouraging. We have thus considered applying (\ref{finalD}) to the realistic Fermi sea, regardless of the validity of this formula for small $k$. Substituting (\ref{finalD}) into (\ref{nlt}), we get another prediction of the relative height of the second plateau to the first one as
\begin{eqnarray}\label{ns}
  \delta n_s  &=& \frac{1}{N} \sum_{k=1}^{k_f} [  \cos ( 2 |l| \eta k - \theta (\eta k )) -\cos (2 |l| \eta k )] - \frac{1}{N} ,  \nonumber  \\
\end{eqnarray}
where the last term corresponds to the state $|0^+\rangle $. We have used the fact that $\theta(0) = \pm \pi $, and that the $k=0$ term has to be scaled down by $1/2$ in view of (\ref{evenstates}). Here the subscript stands for summation. Compared with $\delta n_A $ in (\ref{abelformula}), $\delta n_s $ has the advantage that it is also defined at $l=0$.

The summation above can be rewritten more compactly as $\delta n_s  = (S_1 - S_2 )/(2 N)$, with
\begin{subequations}\label{S1S2}
\begin{eqnarray}
  S_1  &=&  \sum_{k=-k_f}^{k_f}  \cos [2 |l| \eta k - \theta (\eta k )]  ,  \\
  S_2 &=&   \sum_{k=-k_f}^{k_f}  \cos (2 |l| \eta k ) =\frac{\sin (2 q_f |l| + \eta |l| )}{\sin ( \eta |l|)  }  . \quad \quad
\end{eqnarray}
\end{subequations}
The second term was easily calculated as the arguments of the cosine function forms an arithmetic sequence.
The first term is more complicated---Because of the $\theta $ function, the arguments of the cosine function are not an arithmetic sequence. But they are not far away from an arithmetic sequence, especially when $|l| $ is large. Actually, the $\theta$ function defined in (\ref{deftheta}) is a very regular function in the sense that its derivative is bounded. It can be easily verified that $|d\theta /d q |\leq 4/|U|$. Therefore, if $|U|$ or $|l| $ is large enough, the gap between two neighboring arguments is approximately $2 \eta |l| $.

We thus can employ the same idea as for $S_2$, and approximate $S_1$ as
\begin{widetext}
\begin{eqnarray}\label{S1}
  S_1 &=&   \sum_{k=-k_f}^{k_f}   \cos [ 2 |l| \eta k - \theta (\eta k )] \simeq   \sum_{k=-k_f}^{k_f}   \cos \left( 2 |l| \eta k - \frac{1}{2}\left(\theta ( (k +\frac{1}{2})\eta )+ \theta ( (k -\frac{1}{2})\eta )\right ) \right) \nonumber \\
   &=& \sum_{k=-k_f}^{k_f} \frac{\sin [ 2 |l|  (k+\frac{1}{2})\eta - \theta ((k+\frac{1}{2}) \eta  )] -\sin [ 2 |l|  (k-\frac{1}{2})\eta - \theta ( (k-\frac{1}{2})\eta  )] }{2 \sin [\eta |l| -\frac{1}{2}(\theta( (k + \frac{1}{2}) \eta  )- \theta( (k - \frac{1}{2}) \eta  ))]} \nonumber \\
   &\simeq & \sum_{k=-k_f}^{k_f} \frac{\sin ( 2 |l|  (n+\frac{1}{2})\eta - \theta ( (k+\frac{1}{2})\eta  )) -\sin ( 2 |l|  (k-\frac{1}{2})\eta - \theta ( (k-\frac{1}{2})\eta  )) }{2 \sin ( \eta |l| ) } \nonumber \\
   &=& \frac{1}{\sin ( \eta |l| ) }  \sin \left( 2 |l| (k_f+\frac{1}{2}) \eta  - \theta ( (k_f +\frac{1}{2})\eta  ) \right)   \simeq  \frac{1}{\sin ( \eta |l| ) }  \sin ( 2 q_f |l| + \eta |l|  - \theta_f ) .
\end{eqnarray}
\end{widetext}
It is easy to verify that the error introduced in the first and the last line is on the order of $1/N $, and the approximation in the third line is legitimate if $|Ul|  \gg  1 $.

Collecting (\ref{S1S2}) and (\ref{S1}), we see that the formal summation (\ref{ns}) can reduce to (\ref{abelformula}) in the large $l
$ limit. This is verified in Fig.~\ref{fig_abel_sum}. This is an interesting result, as (\ref{abelformula}) and (\ref{ns}) come from the fictitious Fermi sea and the realistic Fermi sea, respectively. That they agree with each other asymptotically adds to the plausibility of both.

\begin{figure}[tb]
\centering
\includegraphics[width= 0.45 \textwidth ]{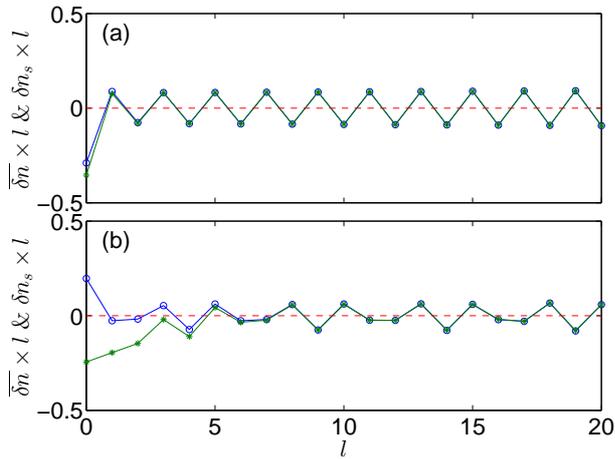}
\caption{(Color online) Comparison between the formal summation (\ref{ns}) (the $\ast$ markers) and the numerical exact result $\overline{\delta n }$ [see Eq.~(\ref{defp}), the $\circ$ markers]. In (a) [(b)], the parameters $(N, k_f, U)$ are of the same values as in Fig.~\ref{fig_time_abel}(a) [Fig.~\ref{fig_time_abel}(b)]. Except for $l= 0 $, we present $\overline{\delta n} \times l $ and $\delta n_s \times l $ instead of $\overline{\delta n }$ and $\delta n_s $.
\label{fig_exact_sum}}
\end{figure}

Now, for $l$ large enough, $\delta n_s $ is close to $\delta n_A$, while the latter in turn is close to $\overline{\delta n }$. Hence, for $l $ large enough, $\delta n_s $ should also be a good approximation or prediction of $\overline{\delta n }$.
In Fig.~\ref{fig_exact_sum}, we compare $\delta n_s $ with $\overline{\delta n }$ for the two sets of parameters in Fig.~\ref{fig_time_abel} or Fig.~\ref{fig_exact_abel}. Indeed, as $l $ increases, the two curves converge together. The two differ significantly only for $l $ small.

\subsection{Influence of the defect mode}\label{subsec_defect}
\begin{figure*}[tb]
\centering
\includegraphics[width= 0.45 \textwidth ]{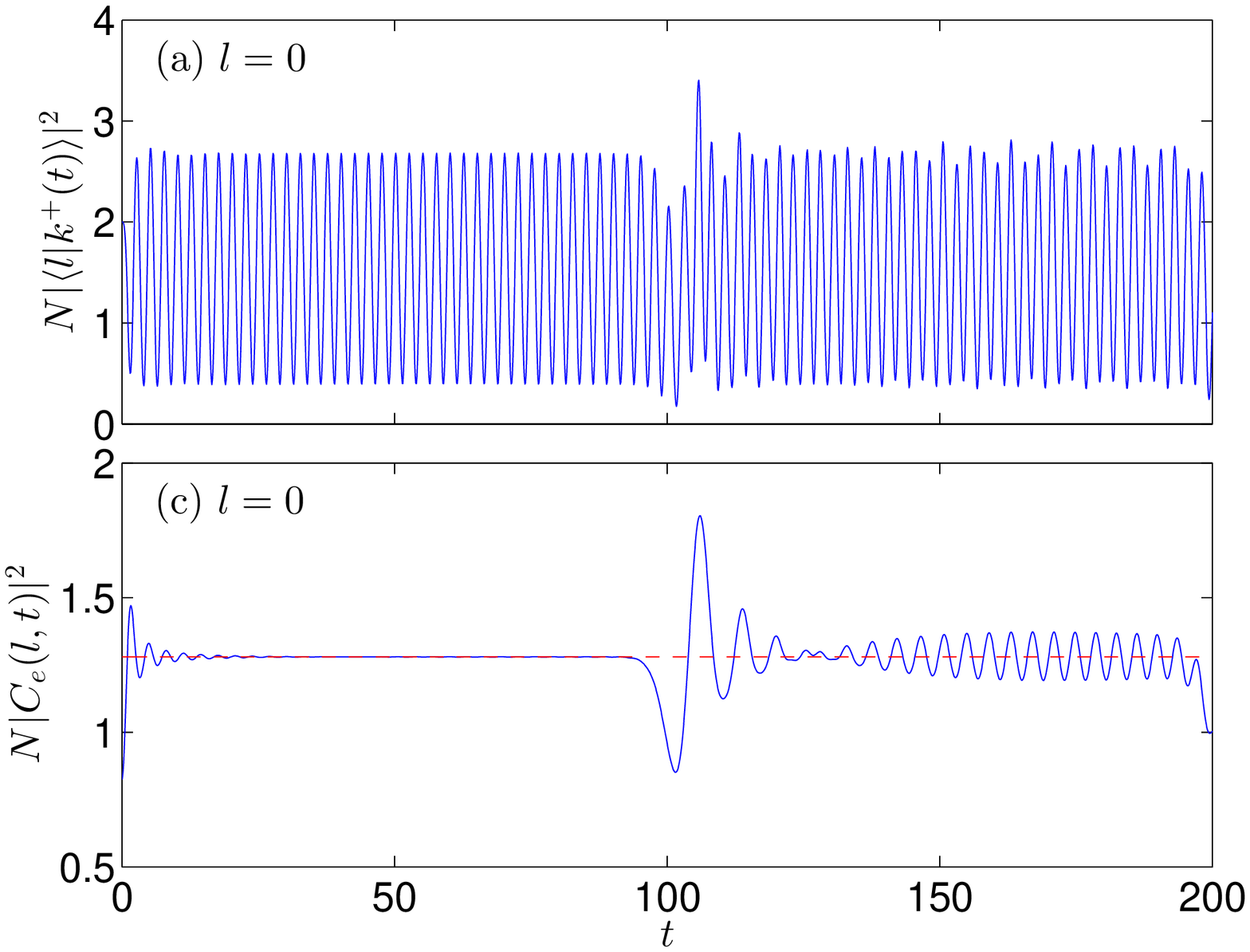}
\includegraphics[width= 0.45 \textwidth ]{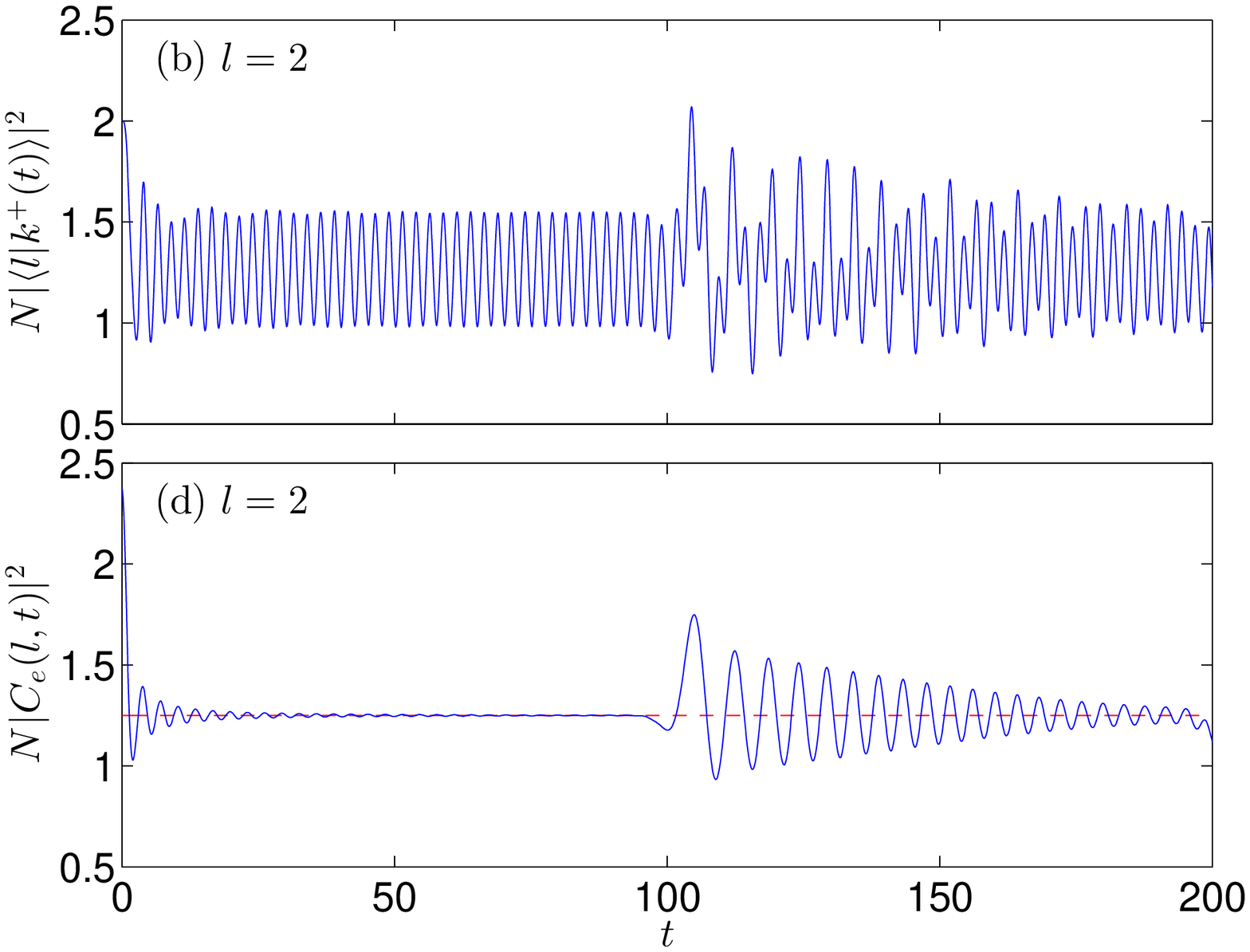}
\caption{(Color online) (a) and (b) Time evolution of the local probability density $|\langle l| k^+(t)\rangle |^2$ for two small values of $l$. The parameters are the same as in Fig.~\ref{fig_standing_standing}(a), i.e., $(N, k, U) = (201, 50, 1.5)$. (c) and (d) Time evolution of the quantity $| C_e |^2$ (see Eq.~(\ref{cet}) for definition). The horizonal dashed line indicates the analytic prediction of (\ref{cet2}).
\label{fig_small_single}}
\end{figure*}

So far, we have obtained two analytic formulas for the height of the second plateau. They both work very well for $l $ large, but become inadequate for $l $ small. The Abel formula (\ref{abelformula}) is not defined at all for $l = 0 $, while the summation formula (\ref{ns}) deviates from the exact value significantly for $l $ around zero.

This local failure forces us to examine the small-$l$ cases more carefully. It turns out that the small $l$'s need extra attention already on the single-particle level. Actually, an attentive reader should have noticed that in Fig.~\ref{fig_snapshots_single}, a spike pulses at $l= 0$, while the pattern of $|\langle l |k_i^+ (t) \rangle |^2$ in its vicinity is mostly stationary. In Figs.~\ref{fig_small_single}(a) and \ref{fig_small_single}(b), the local probability density $|\langle l | k^+(t)\rangle |^2$ as a function of time is shown for $l=0$ and $l=2$, respectively, with all other parameters the same as in Fig.~\ref{fig_standing_standing}(a). We see that unlike in Fig.~\ref{fig_standing_standing}(a), where $l= 50$, for these two small values of $l$, $|\langle l | k^+(t)\rangle |^2$ does not show any sign of plateau, but just oscillates very quickly. More systematic investigation reveals that as $l$ increases, the oscillation amplitude shrinks and the two-plateau structure emerges gradually. The former fact is already visible by comparing Fig.~\ref{fig_small_single}(a) with \ref{fig_small_single}(b).

All these observations point to the defect mode induced by the defect potential. It is well-known that on an infinite one-dimensional tight binding lattice, a defect potential like $H_1$ will induce a defect mode $|\phi_d \rangle $ localized around the defect \cite{feynman}. Its wave function is of the form
\begin{eqnarray}\label{defectmode}
  \langle l | \phi_d \rangle  = \phi_d (l) = A c^{|l|}  ,
\end{eqnarray}
where $c = \frac{1}{2}[U - (\sgn{U} )\sqrt{U^2 + 4}]$. The corresponding eigenenergy is $\epsilon_d = ( \sgn{U} ) \sqrt{U^2 + 4 } $, and the normalization factor $A = \sqrt{U/\epsilon_d }$. In reality, we are dealing with a finite lattice. But as its wave function decays exponentially, as long as the lattice size is much larger than its characteristic size, the defect mode will always be there with its energy and wave function perturbed only slightly.

Besides the defect mode, the post-quench Hamiltonian $H_0 +H_1 $ has also many extended states, whose wave functions spread out on the whole lattice. Let us denote them as $|\phi_m \rangle $ and their energies as $\epsilon_m $ ($1\leq m \leq N - 1 $). Formally, we have
\begin{eqnarray}\label{cet}
  \langle l | k^+(t)\rangle  &=& a_d (k) \phi_d (l)  e^{-i \epsilon_d t } + \sum_{m=1}^{N-1}  a_m(k) \phi_m (l) e^{-i \epsilon_m t } \nonumber \\
  &=& C_d (l,t) + C_e (l,t) ,
\end{eqnarray}
where $a_{d,m}(k)\equiv \langle \phi_{d,m} | k^+ \rangle $, $\phi_{d,m}(l) \equiv \langle l | \phi_{d,m}\rangle $, and in the second line, as the subscripts indicate, $C_d$ and $C_e $ denote the contributions of the bound state and the extended states, respectively.

Now for $l $ large enough, the term $C_d$ is exponentially small and dominated by $C_e$, and $|\langle l | k^+(t)\rangle |^2\simeq |C_e(l,t)|^2$. However, for $l$ small, the term $C_d $ will be competitive
with $C_e $. Although the overlap $a_d $ between the localized mode $|\phi_d \rangle $ and the extended state $|k^+ \rangle$ is on the order of $1/\sqrt{N}$, the amplitude of the mode function at the site $l$, i.e., $\phi_{d}(l) $, is on the order of unity. Hence, the term $C_d $, like $C_e$, is also on the order of $1/\sqrt{N }$.

In Fig.~\ref{fig_standing_standing}(a), it is verified that for appropriate parameters, and when $l$ is large, $|\langle l | k^+(t)\rangle |^2$ shows the predicted plateau structure. As in this region, $\langle l | k^+(t)\rangle \simeq C_e(l, t)$, the observed behavior is actually that of $|C_e(l,t )|^2$. That is, in the time interval $(s_c^+, s_c^-)$,
\begin{eqnarray}\label{cet2}
  |C_e(l,t )|^2 &\simeq & \frac{2}{N} \cos^2 (q |l|  - \theta/2) .
\end{eqnarray}
By analytic continuation, this relation should hold also for $l$ small. This is confirmed in Figs.~\ref{fig_small_single}(c) and \ref{fig_small_single}(d). There, $C_e $ is calculated by subtracting $C_d $ from $\langle l |k^+(t)\rangle $. We see that the well-developed plateau agrees with the prediction of (\ref{cet2}) exactly.

Making use of (\ref{cet2}), and the fact that $C_d$ and $C_e$ contain no common Fourier component, we have on average in the time interval $(s_c^+, s_c^-)$,
\begin{eqnarray}
  \overline{|\langle l | k^+(t)\rangle |^2} &=& \overline{|C_d(l,t)|^2} + \overline{|C_e(l,t)|^2} \nonumber \\
  &\simeq &  |a_d(k)|^2 |\phi_d(l)|^2 + \frac{2}{N} \cos^2 (q |l| - \theta/2) . \quad
\end{eqnarray}
Here the first term on the right hand side is the contribution of the defect mode, which is missed by the ideal model. It can be and was ignored for large $l$. Retrieving this term, we now have a modified prediction of the plateau of $\delta n (l,t)$ as
\begin{eqnarray}\label{dnf}
\delta n_f (l) &=&  P_d  |\phi_d(l)|^2 + \delta n_s (l ) ,
\end{eqnarray}
where $\delta n_s $ is defined in (\ref{ns}) and $P_d$ is the total population on the defect mode, i.e.,
\begin{eqnarray}
  P_d = \sum_{k=0}^{k_f} |a_d(k)|^2 &= & \sum_{k=0}^{k_f} | \langle \phi_d | k^+ \rangle |^2 \nonumber \\
  &= & \sum_{k=-k_f}^{k_f} | \langle \phi_d | k \rangle |^2.
\end{eqnarray}
Note that $P_d$ is $l$ independent.

\begin{figure}[tb]
\centering
\includegraphics[width= 0.45 \textwidth ]{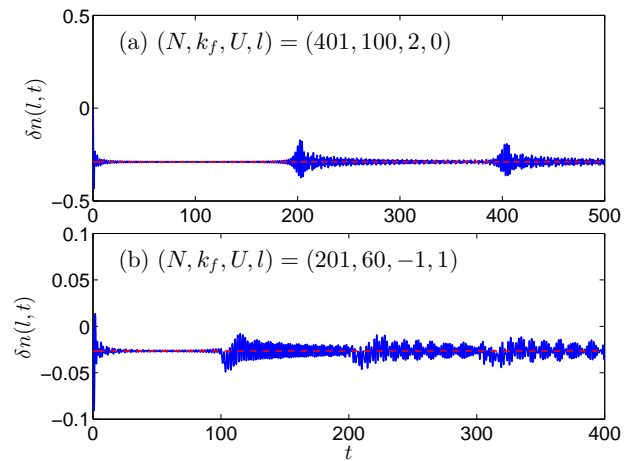}
\caption{(Color online) Time evolution of the variation of the local particle number $\delta n (l,t)$ for two small values of $l$. In each panel, the blue solid line is the numerically exact result, while the horizontal red dashed line is the prediction of Eq.~(\ref{dnf}). The parameters are displayed in each panel.
\label{fig_small_sea}}
\end{figure}

With the contribution of the defect mode taken into account, finally (the subscript of $\delta n_f $ stands for final) we have an accurate formula for the height of the second plateau for all $l$. In Fig.~\ref{fig_exact_sum}, we see that $\delta n_S$ agrees with $\overline{\delta n }$ very well for large $l$, but significantly underestimates it for small $l$. It turns out that the contribution of the defect mode, which is positive-definite, and is peaked at $l=0$, fills the gap exactly. In Fig.~\ref{fig_small_sea}, the time evolution of $\delta n (l,t)$ for two small values of $l$ is shown. We see that the plateaus ensuing the quench are located exactly at the height predicted by $\delta n_f (l)$ in (\ref{dnf}). For large $l$, $\delta n_f $ reduces to $\delta n_s$, and is again accurate.

\section{The infinite lattice case}\label{sec_fo}

\begin{figure}[tb]
\centering
\includegraphics[width= 0.45\textwidth ]{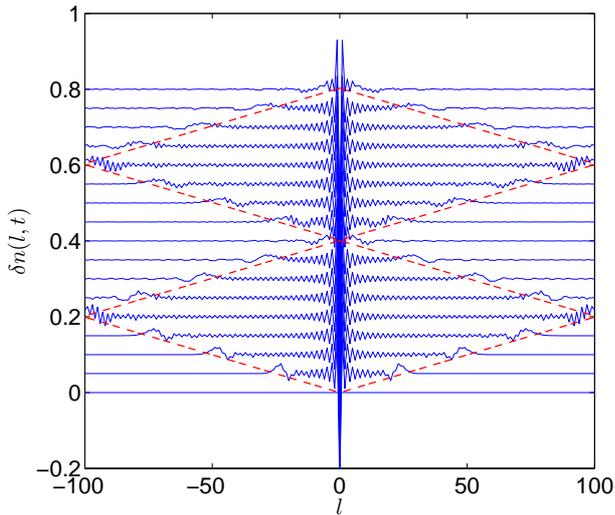}
\caption{(Color online) Equal-distant snapshots of the variation of the local particle number $\delta n (l,t )$ in the time interval $[0, 2T]$. For clarity, two adjacent curves are displaced by $0.05$ in the vertical direction. The parameters are $(N, k_f, U) = (50,25, 2) $. The dashed lines indicate the motion of the wave fronts with a constant velocity of $v_f = 2 \sin q_f \simeq  2 $.
\label{fig_snapshots_sea}}
\end{figure}

\begin{figure*}[tb]
\centering
\includegraphics[width= 0.9\textwidth ]{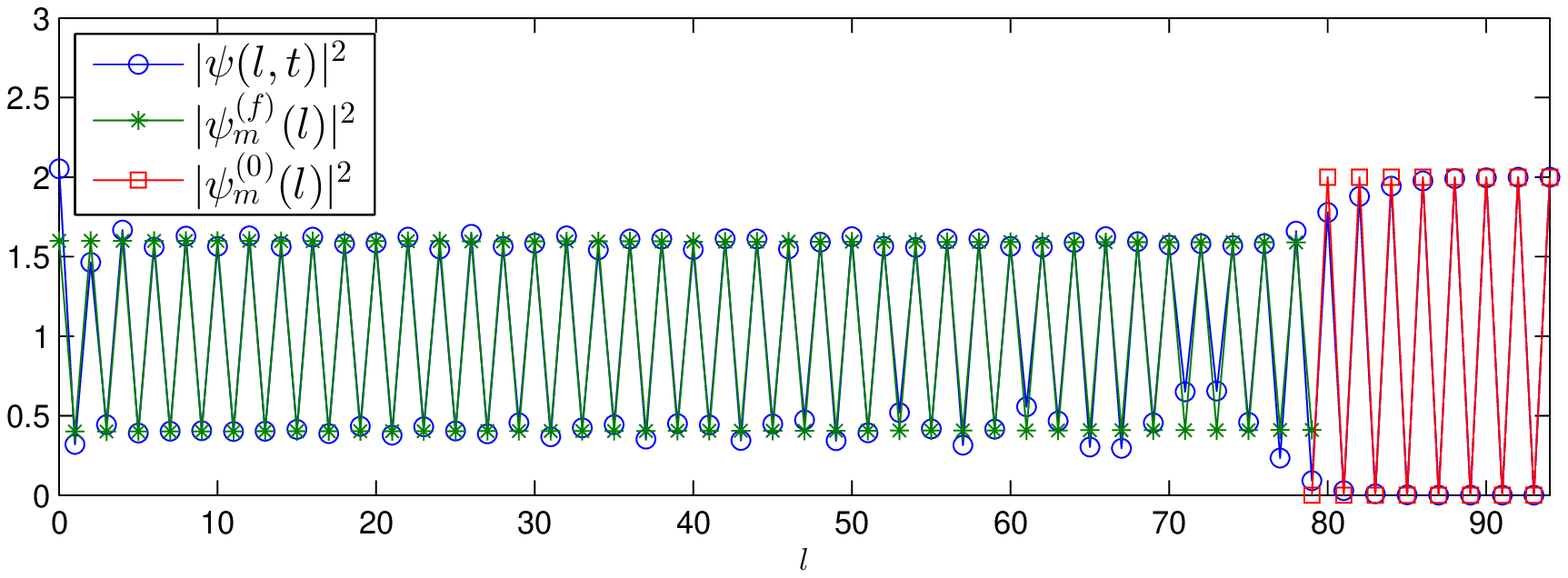}
\caption{(Color online) After the quench, the initial state (\ref{psi0}) relaxes coherently to its adiabatic correspondence (\ref{psif}) inside the light cone, in that the time evolving wave function $\psi(l,t)$ agrees with the eigenstate (\ref{psif}) of the final Hamiltonian inside the light cone, and the eigenstate (\ref{psi0}) of the initial Hamiltonian outside the cone. The parameters are $(N, m, U, t) = (8001, 2000, 1, 40 ) $. Note the size of $N $ here.
\label{fig_crossover}}
\end{figure*}

So far, we have always assumed a finite lattice. On a finite lattice, as we have seen in the figures above, the local particle number $n(l)$ never really settles down to a constant value---again and again, it bursts up suddenly after remaining quiet for a long time. Actually, the out-going waves generated by the sudden quench will sweep across the lattice ring repeatedly, and each time a wave front passes by a site $l$, the local particle number $ \delta n(l)$ experiences an abrupt and violent change.  An overall picture can be well provided by a series of snapshots of $\delta n (l,t)$, as we do in Fig.~\ref{fig_snapshots_sea}. There, like in Fig.~\ref{fig_snapshots_single}, the trajectories of the wave fronts divide the space-time into two different regions, where $\delta n $ show qualitatively different patterns.

But on an infinite lattice, the out-going waves will not come back again. The out-going waves will just pass by each site once and after that the local particle number $n(l)$ will converge to a constant value. That is, the out-going wave fronts leave behind a stationary pattern of $n(l)$, the region of which expands linearly with time. This phenomenon is actually visible in the lower few curves in Fig.~\ref{fig_snapshots_sea}. If superimposed together, one can see that they agree very well in the inner region.

This stationary pattern of $n(l)$ can be fairly called dynamically generated Friedel oscillation.
As discussed in Sec.~\ref{limitation}, as $N\rightarrow \infty$, the ideal model is a good approximation for all $q \in (0,\pi)$. Hence, in this limit, formula (\ref{ns}) and in turn (\ref{dnf}) can be justified almost rigorously. By (\ref{dnf}),  the stationary value of $ \delta n (l, t\rightarrow \infty )$ is
\begin{eqnarray}
  \delta n_{dyn}(l) &=& \delta n_{dyn}^{(d)}(l) + \delta n_{dyn}^{(e)}(l) ,
\end{eqnarray}
where the two terms on the right hand side corresponds to the contribution of the defect mode and the extended states, respectively. By straightforward calculation and replacing summation by integral, we have
\begin{eqnarray}
  \delta n_{dyn}^{(d)}(l) &=& P_d |\phi_d(l)|^2 =   \frac{U^4  c^{2 |l| } }{\pi \epsilon_d^2 }\int_0^{q_f}  \frac{d q }{ (\epsilon_d +  2 \cos q)^2  }\quad \quad
\end{eqnarray}
and
\begin{eqnarray}\label{dndyne}
  \delta n_{dyn}^{(e)}(l) &=&  \frac{1}{2\pi } \int_0^{q_f} [\cos(2 q |l| - \theta) - \cos(2q l )]  dq .
\end{eqnarray}

It is natural to compare the dynamically generated Friedel oscillation with the conventional static (or equilibrium) Friedel oscillation \cite{friedel,harrison, eigler, weismann, giant}. It is well known that a defect in a Fermi sea will induce density oscillation around it. To calculate the density oscillation, a convenient approach is using Green functions \cite{greenfunbook}. But here we shall take a more straightforward way, which will also help us gain some insight of the problem.

Imagine that for the initial Hamiltonian $H_0$, all the Bloch states with wave vectors $|q|\leq q_f $ are filled. Now ramp up the defect potential adiabatically. The odd-parity eigenstates are not affected and their contribution to the particle density is unchanged. But the even-parity eigenstates do feel the potential and deform accordingly. For $H_0$, a generic even-parity eigenstate is of the form
\begin{eqnarray}\label{psi0}
  \psi_m^{(0)}(l) &=& \sqrt{\frac{2}{N}} \cos q  l ,   \quad -L \leq l \leq L ,
\end{eqnarray}
where the wave vector $q$ satisfies $qN = 2 \pi m $ for some integer $m $. For the final Hamiltonian $H_f = H_0 + H_1$, its even-parity eigenstates fall into two categories. The first category contains only one state, i.e., the defect mode, which in the $N \rightarrow \infty $ limit, reduces to $|\phi_d \rangle $ in (\ref{defectmode}). The second category are the extended states. By Bethe's ansatz, their wave functions are in the form
\begin{eqnarray}\label{psif}
  \psi^{(f)}_m(l) &=& B  \cos ( q |l| - \theta/2)  , \quad -L \leq l \leq L ,
\end{eqnarray}
where $\theta \in (-\pi, \pi)$ is some phase shift, and $B$ is a normalization factor, which in the thermodynamic limit should be close to $\sqrt{2/N}$. The eigenvalue equation at $l =0 $ determines the relation between $\theta $ and $q$,
\begin{eqnarray}
  \tan \frac{\theta}{2} &=& \frac{U}{2\sin q } ,
\end{eqnarray}
which is exactly Eq.~(\ref{deftheta}). The only thing is that the $q$'s are to take different values than in (\ref{psi0}). Specifically, the periodic boundary condition requires $\psi^{(f)}(-L) = \psi^{(f)}(L+1 )$, which leads to the condition
\begin{eqnarray}
  q N -  \theta  &=& 2\pi m  ,
\end{eqnarray}
with $m$ being an integer. By continuity, $\psi^{(f)}_m$ is the adiabatic correspondence of $ \psi_m^{(0)} $.

These are interesting results. Comparing (\ref{finalD}), (\ref{psi0}), and (\ref{psif}), it suggests that when we quench an even-parity standing wave like (\ref{psi0}) on a sufficiently large lattice ring, the stationary pattern predicted by (\ref{finalD}) inside the light cone actually corresponds to the eigenstate (\ref{psif}) of the final Hamiltonian. That is, after the abrupt quench, apart from a component proportional to the defect mode, the initial state is phase shifted to its adiabatic correspondence inside the light cone. The resulting probability distribution is largely the same as if the defect potential is turned on adiabatically. This is confirmed in Fig.~\ref{fig_crossover}.

Now we can sum over the even-parity states and determine the static Friedel oscillation. Again, it contains two parts, i.e.,
\begin{eqnarray}
  \delta n_{st} (l) &=& \delta n_{st}^{(d)} (l) +   \delta n_{st}^{(e)} (l),
\end{eqnarray}
where the defect mode's contribution is
\begin{eqnarray}
  \delta n_{st}^{(d)} (l) &=& \begin{cases} |\phi_d(l )|^2  , & \text{if } U < 0 ,\\ 0, & \text{if } U > 0 ,   \end{cases}
\end{eqnarray}
while the extended states' contribution is
\begin{eqnarray}
   \delta n_{st}^{(e)} (l) &=& \int_0^{q_f} \frac{dq}{2 \pi /N } \frac{2}{N } [ \cos^2 (q |l| - \theta /2) - \cos^2 q |l| ] \nonumber \\
   &=&  \frac{1}{2\pi } \int_0^{q_f} \left [\cos(2 q |l| - \theta) - \cos(2q |l| ) \right ] dq  \nonumber\\
   &=&  \delta n_{dyn}^{(e)} (l) .
\end{eqnarray}
That $ \delta n_{st}^{(e)}  =  \delta n_{dyn}^{(e)}  $ results naturally from the single-particle behavior discussed above. Hence, we see that the dynamically generated Friedel oscillation differs from its static counterpart only by an exponentially localized part.

It is instructive to study the asymptotic behavior of $\delta n_{st}(l)$, or that of $\delta n_{dyn}(l)$ for large $|l|$. As both $\delta_{dyn}^{(d)}(l)$ and $\delta_{st}^{(d)}(l) $ decay exponentially, what we want is actually the asymptotic behavior of $ \delta n_{st}^{(e)} (l) $, or equivalently $ \delta n_{dyn}^{(e)} (l) $. The second part of (\ref{dndyne}) can be easily integrated to $\sin (2q_f |l| )/ 4\pi |l| $. As for the first part, it is
\begin{eqnarray}\label{exp3}
  \frac{1}{2\pi } \int_{0}^{q_f} d q \left ( f(q) \cos 2 q |l| +  g(q) \sin 2 q |l| \right )
\end{eqnarray}
with
\begin{eqnarray}
  f(q) &=& \cos  \theta =  \frac{4 \sin^2 q-U^2 }{ 4 \sin^2 q + U^2  } , \nonumber \\
   g(q) &=& \sin  \theta = \frac{4 U \sin q}{4 \sin^2 q + U^2  } . \nonumber
\end{eqnarray}
Both $f$ and $g$ are slowly varying functions of $q$. Integrating by parts, we have
\begin{eqnarray}
& & \frac{1}{2\pi } \int_{0}^{q_f} d q  ( f(q) \cos 2 q |l| +  g(q) \sin 2 q |l| ) \nonumber \\
&=&   \frac{1}{4\pi |l|} [ f (q) \sin 2 q |l|  -  g (q ) \cos 2 q |l|  ] \bigg|_0^{q_f} \nonumber \\
 & & \quad  - \frac{1}{4\pi |l| } \int_{0}^{q_f} d q  ( f'(q) \sin 2 q |l| -  g'(q) \cos 2 q |l| ) \nonumber \\
 & =&    \frac{1}{4 \pi |l|} [ f (q_f) \sin 2 q_f |l|  - g (q_f ) \cos 2 q_f |l|  ] + O \left(\frac{1}{|l |^2} \right ) \nonumber \\
 &=&   \frac{1}{ 4 \pi |l| } \sin (2 q_f |l| -  \theta_f ) + O \left(\frac{1}{|l|^2}\right)  .
\end{eqnarray}
Combing the two parts, we recover the formula (\ref{abellimit}).

\section{Conclusions and discussions}\label{conclusion}

We have investigated the nonequilibrium dynamics of a Fermi sea when a defect potential is introduced suddenly.
On the one hand, this is a natural generalization of our previous study of the quench dynamics of a single Bloch state, which is featured by the cusps or plateaus in the time evolution of some physical quantities. The concern is whether these singular behaviors will survive in the many-particle case. On the other hand, from the point-of-view of the well-known (equilibrium) phenomenon of Friedel oscillation, this is a problem of interest: what if the defect potential is introduced not adiabatically but abruptly?

As for the first problem, the primary observation is that the local particle number $n(l)$ alternates between two plateaus.
This two-plateau behavior is actually exhibited on the single-particle level by an even-parity eigenstate. But its existence on the many-particle level is far from obvious. Let alone other issues, the plateau-switching times for different single-particle states disperse continuously in a wide range, and hence it could be totally reasonable if the abrupt plateau-switching were smeared out.

Leaving  aside the problem of the very existence of the plateaus, we have focused on predicting the heights of the plateaus.  This seems like a mere summation job. But the problem is that we do not really know all the summands, as the theory we had developed for the corresponding single-particle problem holds not for all the  single-particle states involved. We thus had to resort to arguments and approximations, and when confronted with a series oscillatory instead of convergent, we simply took on the Abel summation method to assign a finite value to it. Magically, this finite value is a very accurate prediction of the numerically exact one for $l $ not very small. We are thus in a (possibly fortunate) dilemma that we get the numbers accurate without knowing why. The unreasonable relevance and accuracy of the formalism is yet to be understood. Anyway, the success of the formal approach encouraged us to try other formal expressions, and ultimately an expression accurate for all sites is obtained. We have thus successfully determined the backbone of the evolution trajectory of $n(l)$. This is possibly all that we can wish for, as the erratic details exhibited by $n(l)$ are definitely beyond any simple analytic formula.

As for the second problem, the interesting thing is on the single-particle level. It is found that on an infinite lattice ring, after the quench, the state adjusts itself towards the eigenstate of the final Hamiltonian with equal energy. Specifically, apart from a localized component corresponding to the defect mode, the time evolving state agrees with the phase-shifted eigenstate of the final Hamiltonian inside the light cone. Because of this nice single-particle behavior, the dynamically generated Friedel oscillation differs from the conventional, static Friedel oscillation only by an exponentially decaying term. It is conjectured that this single-particle property holds regardless of the quench potential or the quench protocol, as long as the quench potential is a local one and it stays unchanged in the end.

As stressed as our point, simple models can yield rich dynamics. Here, in the study of a non-interacting many-particle model, we have actually noticed many interesting phenomena which are not accounted for yet. For example, in Fig.~\ref{fig_standing_standing}(c), around $t = 50$, the solid curve oscillates in a very elegant way, which is typical of a state close to the band edge. It reminds us of the Airy function actually. Can we find an analytic formula for it? It might be related to the shape of the out-going  wave fronts. Similarly, in both panels of Fig.~\ref{fig_standing_sea}, in the first period and on the second plateau, the oscillation amplitude shrinks with time. But with what rate does it shrink? Is it power law or exponential?
To answer these questions, likely we have to go beyond the ideal model.
Anyway, more problems are posed than answered, and more effort is to be expended on this simple model.

\section*{Acknowledgments}

We are grateful to J. Guo, R. Huang, K. Yang, H.-T. Yang, and H.-F. Song for helpful discussions. This work is supported by the Fujian Provincial Science Foundation under Grant No. 2016J05004, National Science Foundation of China under Grant No. 11704070, Science Challenge Project under Grant No. JCKY2016212A502, and SPC-Lab Research Fund under Grant No. XKFZ201605.

\appendix*
\section{Derivation of Eq.~(\ref{finalD})}
By (\ref{seq2}), the equation of motion of $\tilde{a}_j(t)$ is
\begin{eqnarray}
  i \frac{\partial}{\partial t} \tilde{a}_j  &=& j \Delta  \tilde{a}_j + 2 g \sum_{m=-\infty}^{\infty} \tilde{a}_m .
\end{eqnarray}
The two terms containing $\Delta$ and $g$ come from $\tilde{H}_0 $ and $\tilde{H}_1$, respectively. The important thing is that the latter is independent of $j$. Hence, we can introduce the auxiliary quantity
\begin{eqnarray}\label{defs}
  S(t) &=& \sum_{m \in \mathbb{Z} } \tilde{a}_m (t) ,
\end{eqnarray}
and solve $\tilde{a}_j (t)$ formally as
\begin{eqnarray}\label{ajt}
  \tilde{a}_j(t) &=& e^{-i  j \Delta t} \delta_{j,0} -i 2g \int_0^t d \tau e^{-i  j \Delta  (t- \tau )} S(\tau) .
\end{eqnarray}
Plugging it into (\ref{defs}), we get an integral equation of $S(t)$,
\begin{eqnarray}
  S(t) &=& 1 - i 2 g \int_0^t d \tau \left(\sum_{m \in \mathbb{Z}} e^{-i  m \Delta (t- \tau)} \right) S(\tau ).
\end{eqnarray}
The integral part is in the form of convolution, which makes it amenable to the Laplace transform. Defining $L(p) = \int_0^\infty d t e^{- p t } S(t)$, we get
\begin{eqnarray}
  L(p) &=& \frac{1}{p } - i 2 g \left(\sum_{m \in \mathbb{Z}} \frac{1}{p+ i  m \Delta } \right ) L(p) .
\end{eqnarray}
Using the Euler formula  $\sum_{m \in \mathbb{Z}} 1/(z + m ) = \pi \cot \pi z $ \cite{book}, we solve $L(p)$ as
\begin{eqnarray}
  L(p) &=&  \frac{1/p}{1 + g T \cot (-i p T/2) } .
\end{eqnarray}
Where $T \equiv 2\pi /\Delta $ is the Heisenberg time associated with the linear spectrum in (\ref{tildeH0}).
This name apparently comes from the time-energy uncertainty relation. As we shall see below, it is the most important time scale in the model, as it determines the period of the curves in Fig.~\ref{fig_standing_sea}.

We want to decompose the meromorphic function $L(p)$ according to its poles and the corresponding residues in the form $L(p) =  \sum_{m \in \mathbb{Z}} b_m/ (p + i E_m )$,
so that we can recover $S(t)$ from $L(p)$ as $S(t) =  \sum_{m\in \mathbb{Z}} b_m e^{-i E_m t } $.
The poles $-i E_m $ of $L(p)$ are roots of the equation $1 + g T \cot (-ip  T/2)  = 0$.
We solve easily
\begin{eqnarray}\label{spectrum}
  E_m  = m \Delta + \gamma  ,
\end{eqnarray}
with $\gamma = \theta (q_i)/T $, where the angle $\theta $ is defined as a function of the wave vector $q$ as
\begin{eqnarray}\label{appendeftheta}
  \theta  \equiv  2 \arctan g T = 2 \arctan \frac{U}{2 \sin q }.
\end{eqnarray}
The corresponding residues are
\begin{eqnarray}
  b_m  &=& \frac{1/p}{g T \frac{d}{dp } \cot (-i p T/2)}\bigg|_{p = - i E_m } = \frac{2g }{1 + g^2 T^2 } \frac{1}{E_m } .\quad \quad
\end{eqnarray}
We thus know that $S(t)$ is of the form
\begin{eqnarray}
  S(t) &=& \frac{2g}{1 + g^2 T^2 }\sum_{m\in \mathbb{Z}} \frac{e^{-i(m\Delta + \gamma )t}}{ m \Delta + \gamma } .
\end{eqnarray}
The summation can be carried out by noting that
\begin{eqnarray}\label{eibx}
  e^{i h x} &=& \frac{e^{i 2 \pi h } -1}{2 \pi i }\sum_{m\in \mathbb{Z}} \frac{e^{im x }}{h - m },\quad 0< x < 2 \pi .
\end{eqnarray}
The result is that, for $t = r T + s$, where $0 < s < T$, and $r$ is a nonnegative integer,
\begin{eqnarray}
  S(t) &=& \frac{e^{-i r \theta }}{ 1 + i g T } ,
\end{eqnarray}
which is a piece-wise constant function. Substituting this result into (\ref{ajt}), we get
\begin{eqnarray}\label{ajt2}
  \tilde{a}_j(t) &=& \left[\delta_{j, 0 } + \frac{2g (e^{-ij \Delta s} -1 )}{\Delta (1+ i g T) j }\right] e^{-i r \theta } .
\end{eqnarray}
Note that the factor in the brackets depends only on $j$ and $s$, while the exponential factor outside depends only on $r$. In the special case of $j =0$, $\tilde{a}_j $ is a piece-wise linear function of $t$. This nonsmoothness results in the cusps in Fig.~\ref{fig_bloch}(a), as $P_{i,r} = |1 \pm  \tilde{a}_0|^2/4$.

Substituting (\ref{ajt2}) into (\ref{approx1}) and ignoring the global phase $e^{-i r \theta - i \varepsilon(q_i)t }$ which is independent of $l$, we get
\begin{eqnarray}\label{approx2}
    \langle l | k_i^+(t)\rangle  & \propto &  \sqrt{\frac{2}{N}}  \bigg[ \cos q_i |l|  \nonumber \\
    & & \quad + \frac{g \left( e^{i q_i |l| } I_+ (s) + e^{-i q_i |l| } I_-(s) \right ) }{(1+i g T )\Delta}    \bigg ],\quad \quad
\end{eqnarray}
where we have introduced the functions ($0< s < T $)
\begin{eqnarray}
  I_\pm (s) &=&  \sum_{m\in \mathbb{Z}} \frac{e^{-i m \Delta s }-1}{m} e^{ \pm i m \eta |l| } .
\end{eqnarray}
We have
\begin{eqnarray}
  I_+(s) &=& - i \Delta s + \sum_{m\neq 0 }\frac{e^{im(\eta |l| - \Delta s )} - e^{i m \eta |l| }}{m } \nonumber \\
   &=& -i \Delta s -\lim_{h\rightarrow 0 } \sum_{m\neq 0 } \frac{e^{im(\eta |l| - \Delta s )} - e^{i m \eta |l| }}{ h- m} \nonumber \\
   &=& -i \Delta s -\lim_{h\rightarrow 0 } \sum_{m\in \mathbb{Z}} \frac{e^{im (\eta |l| - \Delta s )} - e^{i m \eta |l| }}{ h- m } . \quad \quad
\end{eqnarray}
By (\ref{eibx}), we get (note that $|l|< N/2 $)
\begin{eqnarray}\label{Iplus}
  I_+(s) &=& \begin{cases} 0, & \text{if } s \in (0 ,s_c^+ ) , \\ -  2 \pi i, & \text{if } s\in ( s_c^+ ,  T ) ,      \end{cases}
\end{eqnarray}
where $s_c^+ = \eta |l| /\Delta = |l| / v_i $. Similarly,
\begin{eqnarray}\label{Iminus}
  I_-(s) &=& \begin{cases} 0, & \text{if } s \in (0 ,s_c^- ) , \\ -  2 \pi i, & \text{if } s\in ( s_c^- ,  T ) ,      \end{cases}
\end{eqnarray}
where $s_c^- =  \eta (N-|l|) /\Delta = (N-|l|) / v_i  $. Substituting (\ref{Iplus}) and (\ref{Iminus}) into (\ref{approx2}), we get for $-L \leq l \leq L  $,
\begin{eqnarray}\label{appenfinalD}
  D_l^+(t) &\simeq  & \begin{cases} \frac{2}{N} \cos^2 q_i l , & \text{if } s \in (0 ,s_c^+ )  , \\
  \frac{2}{N} \cos^2 (q_i |l | - \theta_i  /2 ) , & \text{if } s \in (s_c^+ ,s_c^- )  , \quad \quad \quad  \\
  \frac{2}{N} \cos^2 q_i l , & \text{if } s\in ( s_c^- ,  T ) ,   \end{cases}
\end{eqnarray}
with
\begin{eqnarray}
  \theta_i  &=& 2 \arctan \frac{U}{2 \sin q_i }.
\end{eqnarray}

In the derivation above, we see that it is the two defining features of the ideal model, i.e., (\ref{tildeH0}) and (\ref{tildeH1}), that are responsible for its exact solvability. We argue that these two nice properties are shared by many other one-dimensional models \cite{scienceopen}. For a generic one-dimensional system, in the midst of its spectrum, both the eigenenergies and the eigenfunctions should be smooth functions of the state index. Hence, at least locally, the two properties should hold and the ideal model could be relevant for the dynamics of the realistic model.

The problem is that its relevance is confined to one dimension. In higher dimensions, the eigenfunctions are indexed with multiple variables (say $k_x$ and $k_y$) and thus generally the eigenenergies are not equally spaced, i.e., (\ref{tildeH0}) does not hold. Therefore, for higher-dimensional generalizations of the local quench scenario considered here, we need new analytic approaches. While this sounds like a challenge, it also means that the dynamic Friedel oscillation in higher dimensions could be fertile of new features.


\end{document}